\documentclass[lettersize,journal]{IEEEtran}
\usepackage{amsmath,amsfonts}
\usepackage{array}
\usepackage[caption=false,font=normalsize,labelfont=sf,textfont=sf]{subfig}
\usepackage{textcomp}
\usepackage{stfloats}
\usepackage{url}
\usepackage{verbatim}
\usepackage{graphicx}
\usepackage{amsthm}
\usepackage{xcolor}
\usepackage[style=ieee, backend=biber,maxnames=99]{biblatex}
\usepackage[colorlinks=true,
            linkcolor=blue,
            citecolor=blue,
            urlcolor=blue]{hyperref}

\usepackage{setspace}
\usepackage{multirow, hhline}

\usepackage[greek,english]{babel}
\usepackage[utf8]{inputenc}
\usepackage{bbm, dsfont}
\usepackage[linesnumbered,ruled,noline, noend]{algorithm2e}

\usepackage{float}

\usepackage[most]{tcolorbox}

\usepackage{threeparttable}

\usepackage{booktabs}

\theoremstyle{thmstyleone}%

\theoremstyle{thmstyletwo}%
\theoremstyle{thmstylethree}%

\newcommand{\ntt}{\operatorname{NTT}}

\newcommand{\rot}{\operatorname{Rot}}

\newcommand{\oners}{\operatorname{rescale}}
\newcommand{\bconv}{\operatorname{BConv}}
\newcommand{\ltt}{\operatorname{LT}}
\newcommand{\moddown}{\operatorname{ModDown}}
\newcommand{\decompose}{\operatorname{Decompose}}

\newcommand{\bitrev}{\operatorname{bitrev}}

\SetKwProg{Fn}{Procedure}{}{} % usage: 

\SetKwFor{ParFor}{parallel for}{do}{}

\begin{document}
\title{Triple-Hoisted Baby-Step Giant-Step Linear Transformation over CKKS Homomorphic Encryption and Hardware Accelerator}
\author{
Sajjad Akherati, and Xinmiao Zhang\\
		\IEEEauthorblockA{
The Ohio State University, Columbus, OH 43210, U.S.
\\
Emails: \{akherati.1, zhang.8952\}@osu.edu
}}
% \thanks{Manuscript created October, 2025; This work was developed by the IEEE Publication Technology Department. This work is distributed under the \LaTeX \ Project Public License (LPPL) ( http://www.latex-project.org/ ) version 1.3. A copy of the LPPL, version 1.3, is included in the base \LaTeX \ documentation of all distributions of \LaTeX \ released 2003/12/01 or later. The opinions expressed here are entirely that of the author. No warranty is expressed or implied. User assumes all risk.}
% }

\markboth{IEEE TRANSACTIONS ON CIRCUITS AND SYSTEMS—I: REGULAR PAPERS ,~Vol.~X, No.~X, February~2026}
{Improved Linear Transformation Algorithms and Hardware Accelerators for CKKS HE Scheme}

\maketitle

\begin{abstract}
Computations can be directly carried out over ciphertexts using homomorphic encryption (HE), which is indispensable for privacy-preserving cloud computing. Linear transformation is widely used in neural networks, including large language models. However, the implementation of linear transformation over HE requires a large number of ciphertext rotations, which incur significant memory and hardware overhead despite existing simplification techniques. This paper proposes a triple-hoisted baby-step giant-step algorithm that decomposes the baby step further to substantially reduce the number of ciphertext rotations needed for the CKKS HE evaluation of linear transformation. Moreover, to reduce off-chip memory access, which contributes to the majority of the latency, a memory-optimized data path is proposed by partitioning the algorithm into multiple phases. Furthermore, an efficient FPGA-based hardware accelerator with an optimized permutation circuit for message routing is designed for the proposed scheme. For a set of typical parameters, the proposed design reduces off-chip memory access by 4.6$\times$ and 2.1$\times$ compared to the best prior accelerator and implementing the best prior algorithm on a hardware platform similar to ours, respectively. Synthesized for Xilinx Virtex UltraScale+ devices, the proposed design achieves a 5.6$\times$ and 2.4$\times$ reduction in overall latency compared with the baseline designs. Furthermore, it reduces LUT, FF, and DSP utilization by 23\%, 74\%, and 19\%, respectively.
\end{abstract}

\begin{IEEEkeywords}
Baby-step giant-step decomposition, CKKS, Hardware accelerator, Homomorphic encryption, Linear transformation.
\end{IEEEkeywords}

\section{Introduction}
% \IEEEPARstart{H}{omomorphic}
Homomorphic encryption (HE) enables computations to be performed directly on encrypted data without requiring decryption. It is essential for realizing machine learning for applications demanding user data privacy \cite{ML_app}, such as medical diagnosis \cite{HE_Medicine_BioDeepHash}, financial data analysis \cite{HE_finance_PPMLT}, and genome sequencing. The Ring Learning With Errors (RLWE) problem is widely adopted in HE schemes, including CKKS \cite{CKKS}, BGV \cite{BGV}, BFV \cite{BFV_RNS_Improved}, and TFHE \cite{TFHE2}. In particular, the CKKS scheme \cite{CKKS} is the most efficient among available HE schemes. In these schemes, a ciphertext consists of two polynomials, $\underline{\mathbf{ct}} = (\mathbf{c}_0, \mathbf{c}_1) \in \mathcal{R}_Q^2$, where $\mathcal{R}_Q = \mathbb{Z}_Q[x]/(x^N + 1)$. The polynomials in $\mathcal{R}_Q$ have degree at most $N-1$ and their coefficients are integers modulo $Q$. $N$ is typically a power of two in the order of thousands and $Q$ has several hundred bits to achieve the desired security level.

The multiplication of long polynomials with large coefficients can be simplified by various approaches. The complexity can be reduced by decomposing the operands and incorporating the modular reduction by $x^N+1$ into the decomposed components \cite{PolyMultJourn}. The number theoretic transform (NTT) \cite{Area-efficient-conflict-free} lowers the complexity by mapping polynomial multiplications to coefficient-wise multiplications in the transformed domain. The hardware acceleration for NTT has been intensively investigated \cite{area-efficient-ntt-HE, Chiu-NTT-TF, Area-efficient-conflict-free, ParhiNTT}. Integer modular multiplication for the coefficients can be optimized using techniques such as Montgomery multiplication, Barrett reduction \cite{MMIDSajjadZhangJSPS, FPGA-mod-mult}, and Karatsuba decomposition. Additionally, the residue number system (RNS) reduces the complexity of integer arithmetic by representing an integer $a\bmod Q$ through its residues with respect to the pairwise co-prime factors of $Q$ \cite{RNSCKKS}. 

Linear transformations (LTs) are widely employed in HE applications such as neural networks \cite{Panther-HECNN}, transformers \cite{FineTuned-LLMHE}, and bootstrapping \cite{Bootstrapping4}. To efficiently utilize the long polynomials in the ciphertexts, multiple input data are packed into the same ciphertext. As a result, ciphertext rotations are needed to shift the entries in the ciphertext polynomials and derive the matrix multiplication result. Ciphertext rotations have very high complexity and need large rotation keys. The HE linear transformation (HE-LT) is carried out through diagonalizing the constant matrix of the LT and packing the entries in the same diagonal into a separate polynomial \cite{Gazelle}. However, it requires a large number of ciphertext rotations for high-dimensional LTs. To mitigate this, the design in \cite{HELibBootstrapping} decomposes the matrix into multiple matrices consisting of different diagonals and applies a baby-step giant-step (BSGS) ciphertext rotation procedure. Additionally, the complexity of ciphertext rotation is reduced by a double hoisting technique \cite{faster_HELib, Bootstrapping4}, which re-orders the computations to enable the combination of intermediate results. 

% Discrete Fourier transforms (DFT) are special LTs that are employed to implement bootstrapping \cite{Bootstrapping4} and secure neural network inference \cite{FALCON_CVPR}. These transformations can be decomposed to sparse matrices with small number of nonzero diagonals, making their complexity lower than that of BSGS approach\cite{cryptoeprint-fast-DFT}. 

Several hardware accelerators over FPGA devices have been developed for HE evaluation of 
LT \cite{APAS, LTHW-Yang-FPGA_CNN, CHAM-HE-Accelerator, HE_MV_BWEfficient, LTHW-FAME} based on the diagonal method in \cite{Gazelle}. The designs in \cite{APAS, LTHW-Yang-FPGA_CNN} target low-dimensional LTs. The CHAM accelerator \cite{CHAM-HE-Accelerator} employs the PackLWE algorithm \cite{PackLWE} to pack the input data without encoding. However, it needs to store a large number of intermediate ciphertexts due to the automorphism operations and the tree structure used for packing intermediate ciphertexts. The design in \cite{HE_MV_BWEfficient} accelerates the diagonal method of LT by exploiting three forms of parallelism: partial-sum parallelism, RNS polynomial parallelism, and coefficient parallelism. Recently, FAME \cite{LTHW-FAME} proposed a hardware accelerator for encrypted matrix-matrix multiplication, which is decomposed and implemented by LTs \cite{Gao2024SecureGEMM}. It improves performance upon \cite{HE_MV_BWEfficient} by combining intermediate results. For large LTs, storing matrices, rotation keys, NTT twiddle factors, and intermediate results may require hundreds of gigabytes of off-chip memory. Consequently, data transfers from off-chip memory introduce substantial latency. To mitigate this overhead, \cite{HE_MV_BWEfficient, LTHW-FAME} reduces off-chip memory access by storing frequently reused data on-chip. Despite these efforts, HE-LT still requires substantial off-chip memory access and has very long latency.

This paper first proposes to decompose the baby step in the BSGS algorithm into two layers to further reduce the number of required ciphertext rotations, each of which mainly consists of Decompose and ModDown steps. Then, hoisting is applied across the three layers of ciphertext rotations, which not only enables the two decomposed layers of the baby step to share Decompose operations, but also reduces the total number of ModDown operations. Both of these modifications lead to significant complexity reduction. By adjusting the algorithm parameters, flexible trade-offs between the number of required rotation keys and computational complexity can be achieved. The second contribution of this work is optimized data paths for the proposed triple-hoisted BSGS (TH-BSGS) algorithm, which substantially reduce off-chip memory access by maximizing the reuse of intermediate results under limited on-chip memory resources. Specifically, our new algorithm is partitioned into phases according to the different types of data dependency. Then computation reordering and partial product accumulation are adopted within each phase to minimize off-chip memory access. The third endeavor is the development of an efficient hardware accelerator for implementing the proposed algorithm. In particular, the complexity of the automorphism used in ciphertext rotation is reduced by half by exploiting data permutation patterns. Our proposed design is evaluated over Xilinx FPGA devices. Compared with the best prior HE hardware accelerator under the same parameter setting and with an implementation of the best prior algorithm on our hardware platform, the proposed design reduces off-chip memory accesses by 4.6$\times$ and 2.1$\times$, respectively, and overall latency by 5.6$\times$ and 2.4$\times$, respectively, for a set of typical HE parameters. In addition, compared to the prior accelerator, the proposed design reduces LUT, FF, and DSP utilization by 23\%, 74\%, and 19\%, respectively.

The organization of this paper is as follows. Section II introduces background information on the CKKS scheme and the BSGS algorithm. Section III presents our new TH-BSGS algorithm, and Section IV proposes the data path optimization. The proposed hardware accelerator is detailed in Section V. Evaluation results and conclusions are provided in Sections VI and VII, respectively.

\section{Background} \label{sec: preliminaries}
This section reviews essential information for the RNS-CKKS scheme \cite{RNSCKKS} and BSGS algorithm for HE-LT.

\subsection{RNS-CKKS Scheme}
In the CKKS scheme \cite{CKKS}, a ciphertext is denoted as $\underline{\mathbf{ct}} = (\mathbf{c_0}, \mathbf{c_1})$, where $\mathbf{c_0}, \mathbf{c_1} \in \mathcal{R}_Q$. To mitigate the complexity of modular operations on the very large polynomial coefficients, the RNS representation factors the modulus $Q$ into $L+1$ pairwise co-prime integers, i.e., $Q = q_0 q_1 \cdots q_L$, where the factors $q_j$ ($0 \leq j \leq L$) have about the same bit lengths and are referred to as the RNS moduli. Accordingly, an integer $a \bmod Q$ can be uniquely expressed as the set $\left\{ a^{(j)} = a \bmod q_j \mid 0 \leq j \leq L \right\}$. As a result, arithmetic operations on $a$, such as multiplication or addition, can be carried out as $L$ independent operations on the corresponding residues $a^{(j)}$ \cite{Combined2}.

In most existing work, such as \cite{HE_MV_BWEfficient, CHAM-HE-Accelerator, APAS, Gazelle, Bootstrapping4}, it is assumed that the constant matrix of the LT is in plaintext and the data input is encrypted. The multiplication of a ciphertext $\underline{\mathbf{ct}} = \{(\mathbf{c}_0^{(j)}, \mathbf{c}_1^{(j)})\}$ and a plaintext polynomial $\mathbf f= \{\mathbf f^{(j)}\}$ ($0\leq j\leq L$) is carried out as 
\begin{align*}
    {\mathbf f} \times \underline{\mathbf{ct}} \!=\! \left\{ \underline{\mathbf c}_\times^{(j)} \!=\! \left(\mathbf c_{\times, 0}^{(j)}, \mathbf c_{\times, 1}^{(j)}\right)\!=\!\left(\mathbf f^{(j)}\mathbf c_0^{(j)}, \mathbf f^{(j)}\mathbf c_1^{(j)}\right) \right\}. %\label{eq: pt-ct mult}
\end{align*}
Random noise vectors have been added during the encryption, and the noise increases after each multiplication. To reduce the noise, the rescaling operation is carried out as
\begin{align}\label{eq: rs}
    \oners(\mathbf c_\times^{(j)}) = \left[q_{L}^{-1}\left(\mathbf c_{\times}^{(j)}-\mathbf c_{\times}^{(L)}\right)\right]_{q_j}, (0\leq j < L),
\end{align}
where $[\cdot]_{q_j}$ denotes modulo reduction by $q_j$. 

In the CKKS scheme \cite{RNSCKKS}, an input vector of $N/2$ entries can be packed into a ciphertext by canonical embedding. A rotation operation \cite{CKKS, RNSCKKS} can be applied to the ciphertext for an input vector to derive another ciphertext corresponding to the same input vector cyclically shifted by $r<N/2$ positions. Let $g$ be an integer that has order $N/2$ modulo $2N$, e.g. $g=5$. The ciphertext rotation by $r$ slots to the direction of the least significant slot can be achieved by first applying the ring automorphism with parameter $g^r \bmod 2N$ as follows:
\begin{align} \label{eq: rotation automorphism function}
\mathbf{\phi}_{r}: \mathbf{c}_l^{(j)}(x) \mapsto \mathbf c_{l}^{(j)}(x^{(g^r \mod 2N)}) \pmod{x^N + 1}.
\end{align}
Denote the secret key of the HE scheme by $\mathbf s(x)$. The result of the above automorphism is a ciphertext encrypted under the transformed secret key $\phi_r(\mathbf s(x)) = \mathbf{s}(x^{(g^r \mod 2N)}) \pmod{x^N + 1}$. Hence, a key switching is required to re-encrypt it under the original secret key $\mathbf{s}(x)$.

\begin{figure}[t]
    \centering
    \includegraphics[width=1\linewidth]{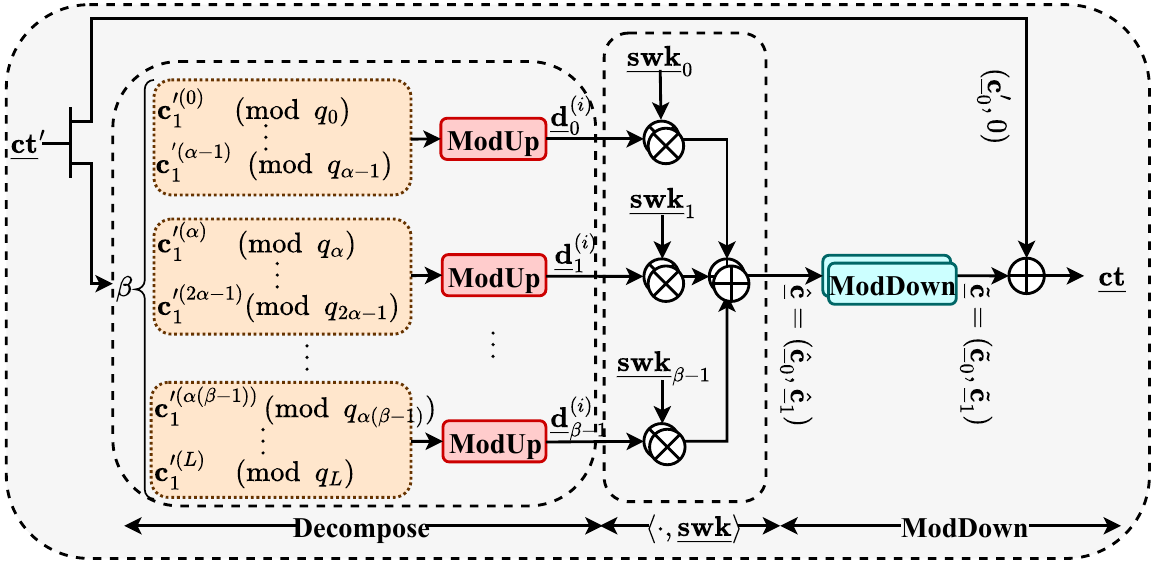}
    \vspace{-15pt}
    \caption{The block diagram for the key switching operation in the RNS-CKKS scheme \cite{Bootstrapping4}.}
    \vspace{0pt}
    \label{fig: key_switching}
\end{figure}

The generalized key-switching procedure in \cite{Bootstrapping3, Bootstrapping4} allows tradeoffs on the computation complexity and multiplicative depth. Its computations are outlined in Fig.~\ref{fig: key_switching}. Let us assume that the ciphertext $\underline{\mathbf{ct}'}={(\mathbf c_0^{\prime(j)}, \mathbf c_1^{\prime(j)})}$ is derived under a secret key $\mathbf s'$. In the key switching, the polynomials $\mathbf c_1^{\prime(j)}$ are firstly divided into $\beta = \lceil (L+1)/\alpha \rceil$ groups, where each group contains $\alpha$ polynomials, except the last group, which may contain fewer than $\alpha$ polynomials. For such a division, a set of $\beta$ switching keys 
$\underline{\mathbf{swk_b}} = (\mathbf{swk_{b,0}}, \mathbf{swk_{b,1}})\in \mathcal{R}_{PQ}^2$ ($0\leq b<\beta)$ are pre-determined based on $\mathbf s'$ and $\mathbf s$ to which switching is performed \cite{Bootstrapping3, Bootstrapping4}. Here $P$ is an integer that can be factored into $\alpha$ co-prime factors, $p_0$, $p_1$, $\cdots$, and $p_{\alpha-1}$, each of which has a similar bit length as $q_l$. Since the modulus of the $b$-th group of the RNS representation of ciphertext $\underline{\mathbf{ct'}}$ is $\prod_{j=\alpha b}^{\min{(\alpha(b+1)-1, L+1)}}q_j$ for $b\in\{0,1,\cdots, \beta-1\}$, a Modup operation needs to be applied to raise its modulus to $PQ$ before it can be multiplied with $\underline{\mathbf{swk_b}}$ as shown in Fig.~\ref{fig: key_switching}. The ModUp is realized by applying a basis conversion process. In particular, a number represented in RNS format with respect to basis $Q$, $\{a^{(j)}\}$, can be converted to the RNS representation with respect to basis $P$ as \cite{RNSCKKS}
\begin{align}
    \bconv (\{a^{(j)}\}, Q, P)  =  \sum_{j=0}^{L}  [\hat{q}_j^{-1} a^{(j)}]_{q_j}\hat{q}_j \pmod{p_i}, \label{eq: basis conversion}
\end{align}
where $\hat{q}_j = Q/q_j$. After the inner product $\hat{\mathbf c} = \langle \underline{\mathbf{d}}, \underline{\mathbf{swk}}\rangle = \sum_{b=0}^{\beta-1}\mathbf d_b\cdot \left(\mathbf{swk}_{b,0}, \mathbf{swk}_{b,1}\right)\mod{PQ}$ is computed, the ModDown operation is carried out to bring the modulus back to $Q$. It is performed as  
\begin{align}\label{eq: moddown}
    \Tilde{\mathbf c}^{(j)} = P^{-1}\left(\hat{\mathbf c}^{(j+\alpha)} - \bconv(\{\hat{\mathbf c}^{(i)}\}, P, Q)\right)\!\!\!\!\mod{\!q_j},
\end{align}
for $0\!\leq\! j \!\leq\! L$ and $0 \!\leq\! i \!<\! \alpha$ \cite{RNSCKKS}. Finally, $\underline{\mathbf{ct}} = (\mathbf c_0^{(j)} \!+\! \Tilde{\mathbf c}^{(j)}_0, \Tilde{\mathbf c}^{(j)}_1)$ is the ciphertext under the secret key $\mathbf s$. The (I)NTT can be applied to reduce the polynomial multiplication complexity. By reformulating the involved computations, the number of (I)NTTs is substantially reduced in \cite{multi-ct-mult}.

\subsection{BSGS Algorithm for HE-LT} 
Without loss of generality, consider an LT whose constant matrix $F$ is of dimension $n\times n$ and $n\leq N/2$. If the matrix size is not a power of two, zeros are padded. If $n > N/2$, the matrix must first be decomposed into smaller blocks with dimensions at most $N/2$, since up to $N/2$ data elements can be packed into a polynomial using the packing scheme in \cite{RNSCKKS}. The diagonal method of HE-LT in \cite{Gazelle} divides $F$ into $n$ diagonal lines. The entries in the $i$-th diagonal are packed into the polynomial $\mathbf{f}_i$ ($0\leq i<n$). Then, the ciphertext corresponding to the linearly transformed input vector can be computed as $\ltt_F(\underline{\mathbf{ct}}) = \sum_{i=0}^{n}\rot(\underline{\mathbf{ct}}, i)\cdot \mathbf{f}_i$, where $\underline{\mathbf{ct}}$ is the ciphertext of the input data vector, and $\rot(\underline{\mathbf{ct}}, i)$ denotes ciphertext rotation by $i$ positions. 

The BSGS algorithm \cite{HELibBootstrapping} improves upon the diagonal method. To reduce the number of ciphertext rotations, which incur significant computational cost as explained in the previous section, the HE-LT is formulated as
\begin{align} \label{eq: bsgs}
    \ltt_F(\underline{\mathbf{ct}}) = \sum_{j=0}^{n_2-1}\rot\left(\sum_{i=0}^{n_1-1}\rot\left(\underline{\mathbf{ct}},i\right)\cdot\hat{\mathbf f}_{n_1j+i}, n_1j\right),
\end{align}
where $n=n_1n_2$ and $\hat{\mathbf f}_{n_1j+i} = \phi_{-n_1j}(\mathbf f_{n_1j+i})$ \cite{HELibBootstrapping}. By decomposing the rotation and summation into two layers as in \eqref{eq: bsgs}, the number of ciphertext rotations needed is reduced from $n-1$ to $n_1 + n_2 - 2$. The complexity can be minimized by setting $n_1 \simeq n_2 \simeq \sqrt{n}$. In \eqref{eq: bsgs}, the inner rotations and summations are referred to as the baby step, while the outer rotations and summations are the giant step.

\SetKwInput{KwParam}{Parameters}
\begin{algorithm}[t]
    \caption{Double hoisted BSGS algorithm (DH-BSGS) for HE-LT\cite{Bootstrapping4}.}  \label{alg: double hoisting bsgs}
    % Inputs Here
        \KwParam{decomposition factors $\alpha$, $\beta$, and $n_1$, $n_2$ such that $n_1n_2=n$.}
        \KwIn{$\underline{\mathbf{ct}}'$, $\widetilde{\underline{\mathbf{swk}}}_{i} = \phi_{-i}(\underline{\mathbf{swk}}_{i})$,  $\widetilde{\underline{\mathbf{swk}}}_{n_1j} = \phi_{-n_1j}(\underline{\mathbf{swk}}_{n_1j})$, $\hat {\mathbf f}_{n_1j+i}\!=\!\phi_{-n_1j}(\mathbf{f}_{n_1j+i})$ ($0\!\leq\! i\!<\!n_1$, $0\!\leq\! j\!<\!n_2$).}
        \BlankLine
        \SetAlgoLined
        
        % Algorithm
        $\underline{\mathbf{d}} \gets \decompose(\underline{\mathbf{c}}'_1, \alpha, \beta, Q, P)$\; 

        $\left(\underline{\mathbf a}_0, \underline{\mathbf b}_0\right)\gets (P\cdot \underline{\mathbf c}'_0, P\cdot \underline{\mathbf c}'_1)\pmod{PQ}$\;
        \For{$i = 0$ \KwTo $i<n_1$}{
            $\underline{\mathbf a}_i\gets \phi_i\left(\underline{\mathbf a}_0 + \langle  \underline{\mathbf{d}}, \widetilde{\underline{\mathbf{swk}}}_{{i,0}}\rangle\right)$\;
            $\underline{\mathbf b}_i\gets \phi_i\left(\langle  \underline{\mathbf{d}}, \widetilde{\underline{\mathbf{swk}}}_{{i,1}}\rangle\right)$\;
        }
        $(\underline{\mathbf c}_0, \underline{\mathbf c}_1)\gets (0,0)$\;

        \For{$j=0$ \KwTo $j<n_2$}{
            $(\underline{\mathbf {u}}_0, \underline{\mathbf u}_1) \gets \sum_{i=0}^{n_1-1}(\underline{\mathbf a}_i, \underline{\mathbf {b}}_i)\cdot \hat{\mathbf {f}}_{n_1j+i}$\;

            $\underline{\mathbf u}_1 \gets \moddown(\underline{\mathbf u}_1)$\;
            $\underline{\mathbf{d}} \gets \decompose(\underline{\mathbf u}_1, \alpha, \beta, Q, P)$\;
            $\underline{\mathbf c}_0 \gets \underline{\mathbf c}_0 + \phi_{n_1j}(\underline{\mathbf u}_0 + \langle \underline{\mathbf{d}}, \widetilde{\underline{\mathbf{swk}}}_{n_1j,0} \rangle)$\;
            $\underline{\mathbf c}_1 \gets \underline{\mathbf c}_1 + \phi_{n_1j}(\langle \underline{\mathbf{d}}, \widetilde{\underline{\mathbf{swk}}}_{n_1j,1} \rangle)$\;
        }
        $\underline{\mathbf{ct}} \gets \oners
        \left(
        \moddown(\underline{\mathbf{c}}_0),
        \moddown(\underline{\mathbf{c}}_1)
        \right)$ \;

        % $\underline{\mathbf{ct}}' \gets \oners(\underline{\mathbf{ct}}')$\;
        \BlankLine
        % Outputs Here
        \KwOut{
            $\underline{\mathbf{ct}}$
        }
\end{algorithm}

The double-hoisted BSGS algorithm \cite{Bootstrapping4} reduces the complexity of the BSGS algorithm, and it is summarized in Algorithm~\ref{alg: double hoisting bsgs}. A ciphertext rotation applies the $\phi$ function as defined in \eqref{eq: rotation automorphism function}, followed by the key switching procedure shown in Fig. \ref{fig: key_switching}, which consists of three steps: Decompose, switching key inner product computation, and ModDown. Since each rotation in \eqref{eq: bsgs} has a different offset, the key switching in each rotation has different inputs and therefore requires a separate Decompose operation. Hoisting delays the application of the $\phi$ function until after the switching key inner product computation, so that all the Decompose operations for the inner layer rotations in \eqref{eq: rotation automorphism function} are performed on the same inputs, requiring only a single Decompose operation. To enable this, the inverse $\phi$ function is applied to the switching keys. The same hoisting is applied to all the rotations in the outer layer of \eqref{eq: rotation automorphism function}. Instead of performing the ModDown operation at the end of each key switching operation, as shown in Fig.~\ref{fig: key_switching}, the DH-BSGS algorithm changes the order of the ModDown operation and the multiplication with $\hat{\mathbf{f}}$ in the inner layer computation of \eqref{eq: bsgs}. As a result, only a single ModDown operation is required on the accumulated sum of the products. Overall, the DH-BSGS algorithm requires $n_2$ Decompose operations, $n_2 + 1$ ModDown operations, and $n_1 + n_2 - 2$ switching keys. Although decreasing $n_2$ reduces the computational complexity, it increases the number of switching keys. Therefore, a trade-off between the memory usage and computational complexity can be achieved by adjusting the parameters.

\section{Proposed Triple-Hoisted BSGS Algorithm} \label{sec: Improved Linear Transformation}
This section proposes a new TH-BSGS algorithm for HE-LT. It decomposes the baby step computation of the original BSGS algorithm into two layers of ciphertext rotations. The hoisting technique is then applied to reduce the Decompose operations. Furthermore, the ModDown operations in all three layers are delayed/combined as much as possible to minimize the computational complexity. With lower computational complexity, the proposed algorithm achieves a substantial reduction in memory requirements compared to the DH-BSGS algorithm.

\begin{algorithm}[t]
    \caption{New TH-BSGS Algorithm for HE-LT.}  \label{alg: Proposed triple hoisted BSGS algorithm}
    % Inputs Here
        \KwParam{decomposition factors $\alpha$, $\beta$, and $n'_1$, $n'_2$, and $n'_3$ such that $n'_1n'_2n'_3=n$.} 
        \KwIn{$\underline{\mathbf{ct'}}$, $\widetilde{\underline{\mathbf{swk}}}_{i}$,  $\widetilde{\underline{\mathbf{swk}}}_{n'_1j}$, $\widetilde{\underline{\mathbf{swk}}}_{n'_1n'_2k}$, and $\hat {\mathbf f}_{n'_1n'_2k+n'_1j+i}\!=\!\phi_{-n'_1n'_2k}(\mathbf{f}_{n'_1n'_2k+n'_1j+i})$ ($0 \leq  i < n'_1$, $0 \leq j < n'_2$, and $0 \leq k<n'_3$),}
        \BlankLine
        \SetAlgoLined
        
        % Algorithm
        $\underline{\mathbf{d}}_0 \gets \decompose(\underline{\mathbf{c}}'_1, \alpha, \beta, Q, P)$\;

        $\left(\underline{\mathbf a}_0, \underline{\mathbf b}_0\right)\gets (P\cdot \underline{\mathbf c}'_0, P\cdot \underline{\mathbf c}'_1)\pmod{PQ}$\;
        \For{$i = 1$ \KwTo $i < n'_1$}{
            $\underline{\mathbf a}_{i}\gets \phi_i(\underline{\mathbf a}_0 + \langle  \underline{\mathbf{d}}_0, \underline{\widetilde{\mathbf{swk}}}_{{i,0}}\rangle)$\;
            $\underline{\mathbf b}_{i} \gets \phi_i(\langle  \underline{\mathbf{d}}_0, \underline{\widetilde{\mathbf{swk}}}_{i,1}\rangle)$\;
            $\underline{\mathbf b}'_{i} \gets \moddown(\underline{\mathbf b}_{i})$\;
            $\underline{\mathbf{d}}_{i}\gets \decompose(\underline{\mathbf b}'_{i}, \alpha, \beta, Q, P)$\;
        } 
        
        \For{$i = 0$ \KwTo $i<n'_1$}{
           \For{$j = 1$ \KwTo $j<n'_2$}{
                $\underline{\mathbf a}_{n'_1j+i} \gets \phi_{n'_1j}(\underline{\mathbf a}_{i} + \langle \underline{\mathbf{d}}_i, \underline{\widetilde{\mathbf{swk}}}_{n'_1j,0} \rangle)$\;
                $\underline{\mathbf{b}}_{n'_1j+i} \gets \phi_{n'_1j}(\langle \underline{\mathbf{d}}_i, \underline{\widetilde{\mathbf{swk}}}_{n'_1j,1} \rangle)$\;
            } 
        }
        $(\underline{\mathbf c}_0, \underline{\mathbf c}_1)\gets \sum_{i=0}^{n'_1n'_2}(\underline{\mathbf a}_i, \underline{\mathbf b}_i) \cdot \underline{\hat{\mathbf f}}_{i}$\;
        \For{$k=1$ \KwTo $k<n'_3$}{
            $(\underline{\mathbf{u}}_0, \underline{\mathbf{u}}_1) \gets \sum_{i=0}^{n'_1n'_2}(\underline{\mathbf a}_i, \underline{\mathbf b}_i) \cdot \underline{\hat{\mathbf f}}_{n'_1n'_2k+i}$\;
            $\underline{\mathbf{u}}_1 \gets \moddown (\underline{\mathbf{u}}_1)$\;
            $\underline{\mathbf{d}} \gets \decompose(\underline{\mathbf u}_1, \alpha, \beta, Q, P)$\;
            $\underline{\mathbf c}_0 \gets \underline{\mathbf c}_0 + \phi_{n'_1n'_2k}(\underline{\mathbf u}_0 + \langle \underline{\mathbf{d}}, \underline{\widetilde{\mathbf{swk}}}_{{n'_1n'_2k,0}} \rangle)$\;
            $\underline{\mathbf{c}}_1 \gets \underline{\mathbf{c}}_1 + \phi_{n'_1n'_2k}(\langle \underline{\mathbf{d}}, \underline{\widetilde{\mathbf{swk}}}_{{n'_1n'_2k,1}} \rangle)$\;
        }
        $\underline{\mathbf{ct}} \gets \oners
        \left(
        \moddown(\underline{\mathbf{c}}_0),
        \moddown(\underline{\mathbf{c}}_1)
        \right)$ \;

        % $\underline{\mathbf{ct}}' \gets \oners(\underline{\mathbf{ct}}')$\;
        \KwOut{
            $\underline{\mathbf{ct}}$
        }
\end{algorithm}

% \onecolumn
% \begin{equation}
%     \ltt_F(\underline{\mathbf{ct}}) &= \sum_{k=0}^{n'_3-1}\rot\bigg(\big(\sum_{j=0}^{n'_2-1}\sum_{i=0}^{n'_1-1}\rot\big(\rot(\underline{\mathbf{ct}},i),n'_1j\big)\quad \cdot \hat{\mathbf f}_{n'_1n'_2k+n'_1j+i}\big), n'_1n'_2k\bigg). \label{eq: bsgs new}
% \end{equation}
% \twocolumn

Decompose $n$ into $n'_1 n'_2 n'_3$. Then ~\eqref{eq: bsgs} can be reformulated as
\begin{align} 
    \ltt_F(\underline{\mathbf{ct}}) &= \sum_{k=0}^{n'_3-1}\rot\bigg(\big(\sum_{j=0}^{n'_2-1}\sum_{i=0}^{n'_1-1}\rot\big(\rot(\underline{\mathbf{ct}},i),n'_1j\big) \nonumber \\ 
    &\quad \cdot \hat{\mathbf f}_{n'_1n'_2k+n'_1j+i}\big), n'_1n'_2k\bigg). \label{eq: bsgs new}
\end{align}
%The above reformulation requires $n_1'-1 + (n'_2-1)n'_1 + n'_3-1 =  n'_1 n'_2 + n'_3 - 2$ rotations. 

By applying hoisting and other reformulations to eliminate and/or combine computations in this formula, our new TH-BSGS algorithm is proposed, as summarized in Algorithm~\ref{alg: Proposed triple hoisted BSGS algorithm}. By applying the hoisting technique, a single Decompose operation can be shared among all inner rotations for $1 \le i < n'_1$. This is reflected in Line 1 of Algorithm~\ref{alg: Proposed triple hoisted BSGS algorithm}. Since the $\underline{\mathbf{a}}_i$ values in Line 4 of Algorithm~\ref{alg: Proposed triple hoisted BSGS algorithm} are not multiplied by any switching keys, the corresponding ModDown operations can be delayed until the end. However, the inner-layer rotations are immediately followed by the middle-layer rotations rather than by sum-of-products computations. Therefore, the ModDown operation on $\underline{\mathbf{b}}_{n'_1j+i}$ cannot be delayed in the same manner as in the DH-BSGS algorithm. For the middle layer of rotations in \eqref{eq: bsgs new}, the hoisting technique is also applied, requiring only $n'_1 - 1$ rather than $n'_1(n'_2-1)$ Decompose operations, as shown in Line 7 of Algorithm~\ref{alg: Proposed triple hoisted BSGS algorithm}. Since this stage is followed by sum-of-products computations, the ModDown operations can be delayed and applied only to the accumulated sum, as shown in Line 15. This reduces the number of ModDown operations from $n'_1(n'_2-1)$ to $n'_3-1$. Similarly, the hoisting technique can be applied to the outer layer of rotations, and the ModDown operations are applied to the sum of the results of the outer rotation layer as shown in Lines 17--19. In addition, the rescale operations can be combined with the ModDown operation in Line 19 to further reduce the overall complexity \cite{Castro2021DoesFH, MAD}.

% \begin{figure}
%     \centering
%     \includegraphics[width=0.95\linewidth]{figures/tripleHistedComplexityAnalysis.pdf}
%     \caption{Number of modular multiplications and switching key size of different LT algorithms for $n=2^{15}$, $L=31$, $w=54$, and $\alpha=12$.}
%     \label{fig: tripleHisted_ComplexityAnalysis}
% \end{figure}

\renewcommand{\arraystretch}{1.4}
\begin{table*}[t]
\caption{Complexity comparisons of TH-BSGS algorithm and prior methods for  ring dimension $N$, matrix dimension $n \leq N/2$, decomposition parameters $\alpha$ and $\beta$, $L+1$ RNS moduli of bit length $w$, and $n'_1 n'_2 n'_3 = n_1 n_2 = n$.}
\label{tab: triple hoisted BSGS complexity analysis}
\centering
\setlength{\tabcolsep}{6pt} % default is 6pt
\footnotesize
\begin{tabular}{c|cccc}
\toprule
\textbf{Algorithm} & \textbf{Decompose} & \textbf{ModDown} & \textbf{Coeff.-Wise Poly. Mult.} & \textbf{\# of Switching Keys} \\
\hline
Diagonal method \cite{Gazelle} 
&  
$1$
& 
$2$
&  
\begin{tabular}{c}
    $2\beta(n - 1)(L + 1 + \alpha)$ \\
    $+ 2n (L + 1 + \alpha)$
\end{tabular}
&
\begin{tabular}{c}    
  $\beta(n  -  1)(L  +  1  +  \alpha)$ \\
\end{tabular}
\\  \hline
BSGS \cite{HELibALG}  
&  
$n_1+n_2-2$
& 
$2(n_1+n_2-2)$
& 
\begin{tabular}{c}
    $2\beta(n_1  +  n_2  -  2)(L  +  1  +  \alpha)$ \\
    $+ 2n (L + 1)$
\end{tabular}
&   
\begin{tabular}{c}
    $\beta(n_1  +  n_2  -  2)(L  +  1  +  \alpha)$ 
\end{tabular}
\\ \hline
\begin{tabular}{c}
    DH-BSGS \cite{Bootstrapping4} 
\end{tabular}  
&  
$n_2$
&  
$n_2 + 1$
&  
\begin{tabular}{c}
    $2\beta(n_1  +  n_2  -  2)(L  +  1  +  \alpha)$ \\
    $+ 2n (L + 1 + \alpha)$
\end{tabular}
&   
$\beta(n_1  +  n_2  -  2)(L  +  1  +  \alpha)$
\\ \hline
\begin{tabular}{c}
    Proposed TH-BSGS 
\end{tabular}
&  
$n'_1 + n'_3$
&  
$n'_1 + n'_3 + 1$
&  
\begin{tabular}{c}
    $2\beta(n'_1 n'_2  \! + \! n'_3 \! - \! 2)( L \! + \! 1 \! + \! \alpha )$ \\
    $+ 2n (L + 1 + \alpha)$
\end{tabular}
&  
$\beta(n'_1 \! + \! n'_2  \! + \! n'_3 \! - \! 3)( L \! + \! 1 \! + \! \alpha )$
\\ 
\bottomrule
\end{tabular}
\end{table*}

\begin{table}[]
    \centering
    \caption{Number of modular multiplications and switching key size of different LT algorithms for $n=2^{15}$, $L=31$, $w=54$, and $\alpha=12$.}
    \label{tab: tripleHisted_ComplexityAnalysis}
    \setlength{\tabcolsep}{1.2pt} % default is 6pt
    \footnotesize
    \begin{tabular}{cc|ccc}
        \toprule
         \textbf{Algorithm} & \textbf{} & \textbf{Param.} & \textbf{Mod. Mul.} & \textbf{SWK Mem. (GB)}  \\ \hline
         BSGS \cite{HELibALG} & - & ($2^8,2^7$) & $2.74\times 10^{11}$ & 41.55  \\ \hline
         DH-BSGS \cite{Bootstrapping4} & \begin{tabular}{c}
              Min. Mem.  \\
              Min. Comp.
         \end{tabular}
         & 
         \begin{tabular}{c}
              ($2^8,2^7$)  \\
              ($2^9,2^6$) 
         \end{tabular}
         & 
         \begin{tabular}{c}
              $2.21\times 10^{11}$  \\
              $2.11\times 10^{11}$ 
         \end{tabular}
         & 
         \begin{tabular}{c}
              41.55  \\
              62.43 
         \end{tabular}
         \\ \hline
         TH-BSGS & \begin{tabular}{c}
              Min. Mem.  \\
              Min. Comp.
         \end{tabular}
         & 
         \begin{tabular}{c}
              ($2^5, 2^5, 2^5$)  \\
              ($2^2,2^7,2^6$)
         \end{tabular}
         & 
         \begin{tabular}{c}
              $2.20\times 10^{11}$  \\
              $2.12\times 10^{11}$ 
         \end{tabular}
         & 
         \begin{tabular}{c}
              10.11  \\
              20.99 
         \end{tabular}
         \\
         \bottomrule
    \end{tabular}
\end{table}

Table~\ref{tab: triple hoisted BSGS complexity analysis} compares the complexity of the proposed TH-BSGS method with the diagonal method \cite{LTHW-FAME}, the BSGS algorithm \cite{HELibALG}, and the double-hoisted BSGS algorithm \cite{Bootstrapping4}. The computational complexity is mainly dominated by Decompose, ModDown, and coefficient-wise polynomial multiplication operations. Compared with the switching keys, the twiddle factors required for the (I)NTTs in the Decompose and ModDown blocks are very small; therefore, they are omitted from the comparison. From Algorithm~\ref{alg: Proposed triple hoisted BSGS algorithm}, it can be directly observed that the TH-BSGS algorithm requires $n'_1 + n'_3 - 1$ Decompose operations, $n'_1 + n'_3$ ModDown operations, and $n'_1 + n'_2 + n'_3 - 3$ switching keys. Since each switching key has modulus $PQ$, it is represented using $(L+1+\alpha)$ RNS components. Following the notations in \cite{LTHW-FAME}, the RNS components are also referred to as limbs in this paper. In addition, each switching key consists of two polynomials. The multiplications with the switching keys contribute to the first term of the coefficient-wise polynomial multiplication complexity listed in Table~\ref{tab: triple hoisted BSGS complexity analysis}. The second term arises from multiplying the input ciphertext by the diagonals of the LT constant matrix. Since the BSGS algorithm carries out these computations over modulus $Q$, rather than modulus $PQ$ as in the other three methods listed in Table~\ref{tab: triple hoisted BSGS complexity analysis}, its multiplications are performed over $L+1$ limbs instead of $L+1+\alpha$ limbs.

In the proposed design, trade-offs between computational complexity and the number of switching keys can be achieved by tuning $n'_1$, $n'_2$, and $n'_3$, subject to the constraint $n'_1 n'_2 n'_3 = n$. Assume that $w = 54$ bits are used to represent each polynomial coefficient in RNS format. For an example case of $N = 2^{16}$, $n = 2^{15}$, $\alpha = 12$, and $L = 31$, Table \ref{tab: tripleHisted_ComplexityAnalysis} presents the complexities of the proposed design in two settings: one with parameters chosen to minimize switching key size and the other for minimizing computation complexity, which is estimated using the total number of coefficient modular multiplications from Decompose, ModDown, and polynomial multiplications. It can be observed that the proposed design has much smaller switching key size but similar computation complexity compared to the DH-BSGS scheme. The computation complexities of different settings are similar in the DH-BSGS and TH-BSGS designs because they are mainly contributed by the polynomial multiplication. The switching key size of the proposed TH-BSGS design is minimized to $\mathcal{O}(\sqrt[3]{n})$ when $n'_1 \simeq n'_2 \simeq n'_3 \simeq \sqrt[3]{n}$. On the other hand, the minimum switching key size of DH-BSGS algorithm is $\mathcal{O}(\sqrt{n})$. The diagonal method requires a very large number of polynomial multiplications and switching keys and is therefore omitted from further comparison.

\section{Memory-Optimized HE-LT Datapath} \label{sec: memory-optimized MVM datapath}
The software implementation library \cite{lattigoLibrary} developed with the DH-BSGS algorithm \cite{Bootstrapping4} carries out the computations in Algorithm 1 step by step in a straightforward manner. It needs to access $>$200 GB off-chip data to evaluate the LT during bootstrapping for HE with typical parameters. Accessing off-chip memory introduces significant latency and accounts for a large part of the overall HE-LT latency. Although the proposed TH-BSGS algorithm substantially reduces the number of required switching keys, the intermediate data access contributes to the majority of off-chip memory access. As a result, the TH-BSGS algorithm also has the same bottleneck of long memory access latency if it is implemented according to Algorithm 2 line by line. 

In this section, a memory-optimized data path is proposed to maximize on-chip data reuse and minimize off-chip memory traffic for the TH-BSGS algorithm. Although data path optimizations have previously been proposed for the diagonal method \cite{LTHW-FAME, HE_MV_BWEfficient}, the hoisting technique along with the BSGS method changes the order of computations. Therefore, the data path must be redesigned accordingly.

The proposed TH-BSGS algorithm exhibits three types of data dependency across all involved computations: I) the (I)NTT operations require all $N$ coefficients of a polynomial limb, such as in Line 5 of Algorithm \ref{alg: Proposed triple hoisted BSGS algorithm}; II) the basis conversion operations require all limbs of a polynomial to perform the Decompose and ModDown operations, such as in Line 6 of Algorithm \ref{alg: Proposed triple hoisted BSGS algorithm}; and III) the rotations with different indices can be carried out simultaneously. One example is the loops formed by Line 8--11. During the execution of Algorithm~\ref{alg: Proposed triple hoisted BSGS algorithm}, when the data dependency changes from Type I to Type II, all limbs resulting from the Type I computation must be stored before the Type II computation is carried out. These intermediate results are large and need to be stored in off-chip memories. Similarly, when the loops cover the rotation indices in different orders, intermediate results also need to be stored off-chip. Algorithm~\ref{alg: Proposed triple hoisted BSGS algorithm} can be partitioned into six phases as follows whenever off-chip memories are needed to store intermediate results.
Phase 1 consists of Lines 1--5 of Algorithm~\ref{alg: Proposed triple hoisted BSGS algorithm}. Since the ModDown operation in Line 6 requires all RNS components before it can begin, the ModDown and Decompose operations in Lines 6 and 7 are grouped into Phase 2. The two nested `for' loops in Lines 8--11 and the computation in Lines 12--14 access data using different patterns. Therefore, they are separated into Phases 3 and 4, respectively. Since each limb generated by the ModDown and Decompose operations in Lines 15--16 is immediately consumed by the computations in Lines 17--18, these operations are grouped into Phase 5. Finally, Phase 6 consists of the final ModDown and rescaling operations in Line 19. By reordering the computation within different phases and applying the limb-level fusion, the intermediate results are reused as much as possible under the limited on-chip memory resources. 

\begin{figure*}[t]
    \centering
    \includegraphics[width=1\linewidth]{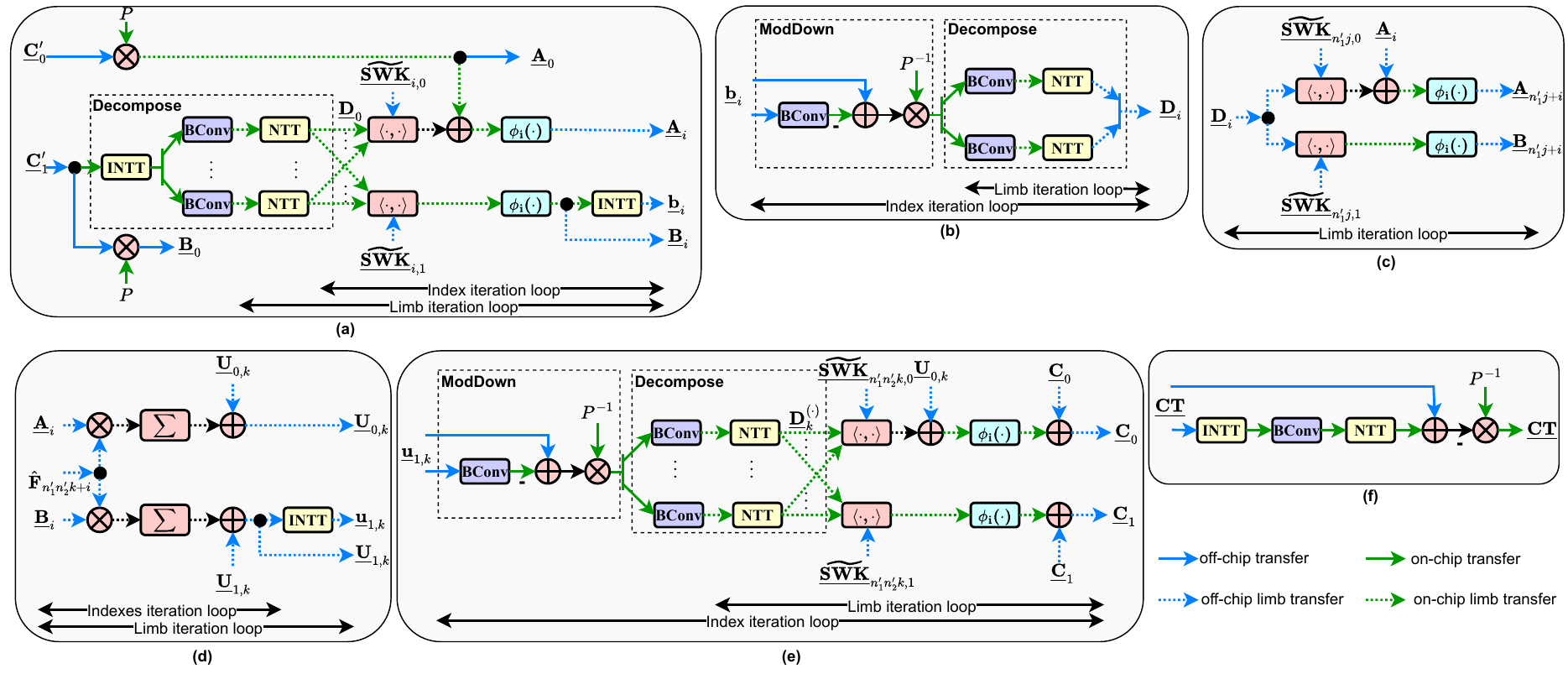}
    \vspace{-15pt}
    \caption{Block diagrams of the proposed memory-optimized data path of (a) Lines 1--5 (Phase 1); (b) Lines 6--7 (Phase 2); (c) Lines 8--11 (Phase 3); (d) Line 12--14 (Phase 4); (e) Lines 15--18 (Phase 5); (f) Line 19 (Phase 6) of the TH-BSGS Algorithm  \ref{alg: Proposed triple hoisted BSGS algorithm}.}
    \label{fig: MO TH_BSGS BD}
    \vspace{-10pt}
\end{figure*}

The data flow for each phase is illustrated in Fig.~\ref{fig: MO TH_BSGS BD}. In this figure, an NTT operation is applied before each polynomial multiplication, and an INTT operation is applied to transform the results back to the polynomial domain for modulus switching. The (I)NTT blocks are explicitly shown in the figure, and the polynomials expressed in the NTT domain are denoted using uppercase boldface letters (e.g., $\mathbf{A} = \ntt(\mathbf{a})$). The blue and green lines in Fig.~\ref{fig: MO TH_BSGS BD} represent reading/writing from/to off-chip and on-chip memories, respectively. The black lines indicate data dependencies without memory accesses. To reduce the latency, ping-pong buffers are employed for reading/writing data from/to off-chip memory. The solid lines in Fig.~\ref{fig: MO TH_BSGS BD} denote simultaneous reading/writing of all polynomial limbs from/to memory.  Due to the limited on-chip memory, some functions process only a subset of the limbs at a time; these operations are represented by dashed lines in Fig.~\ref{fig: MO TH_BSGS BD}. The data flow in the proposed design has been optimized to maximize data reuse and minimize off-chip memory access as detailed below.

\textbf{Phase 1}: The computations in Lines 1--5 of Algorithm~\ref{alg: Proposed triple hoisted BSGS algorithm} are performed in this phase, and the corresponding data flow is illustrated in Fig.~\ref{fig: MO TH_BSGS BD}(a). Since the input to the Decompose block contains only $L+1$ limbs, the INTT operations for all components are carried out in parallel, and the results are stored in on-chip memory. However, the output of each BConv block contains $L+1+\alpha$ limbs, and the available on-chip memory may not be sufficient to store all of these results simultaneously. In our proposed design, $l_1$ ($l_1 \leq L+1+\alpha$) limbs are computed at a time for each of the $\beta$ BConv units, requiring $\left\lceil (L+1+\alpha)/l_1 \right\rceil$ rounds to complete the computation. The NTT computation then follows, and each limb is simultaneously multiplied by $m_1$ switching keys. It takes $\left\lceil (n_1-1)/m_1 \right\rceil$ iterations to complete the multiplication with all switching keys associated with the $l_1$ limbs. Therefore, the overall number of iterations in this process is $\left\lceil (L+1+\alpha)/l_1 \right\rceil \left\lceil (n_1-1)/m_1 \right\rceil$. %The parameters $l_1$ and $m_1$ can be tuned to achieve a memory-computation complexity trade-off.

Instead of generating all the $L+1+\alpha$ limbs in Line 1 of Algorithm \ref{alg: Proposed triple hoisted BSGS algorithm} and then moving to Lines 2-5, our optimized data path finishes the rest of the computations for Phase 1 for the $l_1$ limbs generated by each BConv before continuing the BConv for the other limbs. This allows the generated $l_1$ limbs and the twiddle factors to be reused for the rotations and in the last-stage INTT shown in Fig. \ref{fig: MO TH_BSGS BD}(a). For Phase 1, the following values need to be loaded from off-chip memory: the $2(L+1)$ limbs of the input polynomials $\underline {\mathbf {CT}}'$, the $(L+1)$ limbs of the twiddle factors for the first INTT in Fig. \ref{fig: MO TH_BSGS BD} (a), $(L+1+\alpha)$ limbs each for the twiddle factors for the NTTs in the middle and INTT at the end, and $2(n'_1-1)\beta(L+1+\alpha)$ limbs for the switching keys. The data to be written to off-chip memory includes $L+1$ limbs each for $\underline{\mathbf{A}}_0$ and $\underline{\mathbf{B}}_0$, and $(n'_1-1)(L+1+\alpha)$ limbs each for $\underline{\mathbf{A}}_i$, $\underline{\mathbf{B}}_i$ and $\underline{\mathbf{b}}_i$. On-chip memory is needed to implement input/output buffers and store intermediate computation results. To minimize the peak on-chip memory requirement, once a value is no longer needed for later computations, the memory is re-used to store other values. On-chip ping-pong buffers are used to store the twiddle factors of (I)NTTs, switching keys, $\underline{\mathbf{A}}_i$, $\underline{\mathbf{B}}_i$, and $\underline{\mathbf{b}}_i$. Since the buffer for the first INTT can be re-used for the NTTs in the middle of Fig. \ref{fig: MO TH_BSGS BD}, the sizes of these buffers are $4l_1$, $4\beta m_1l_1$, $2m_1l_1$, $2m_1l_1$, and $2m_1l_1$ limbs, respectively. The $(L+1)$ limbs of $\underline{\mathbf{A}}_0$ need to be stored in the on-chip memory until they are utilized. Although the memory holding the $(L+1)$ limbs of $\underline{\mathbf{C}}_1'$ can be re-used to store the output of the first INTT, the BConv units carry out computations in rounds and require additional memory to store the $\beta l_1$ limbs they generated at a time.

Table~\ref{tab: MO_TH_BSGS complexity} summarizes the breakdown of off-chip memory access and peak on-chip memory requirement for the proposed memory-optimized TH-BSGS algorithm across different phases in terms of the number of limbs. %The overall on-chip memory requirement is determined by the maximum value among all six phases.

\renewcommand{\arraystretch}{1.4}
\begin{table*}[t]
\caption{Memory requirement for the proposed optimized data path of the TH-BSGS algorithm.}
\label{tab: MO_TH_BSGS complexity}
\centering
\setlength{\tabcolsep}{4pt} % default is 6pt
\footnotesize
\begin{tabular}{c|ccccc|c}
\toprule
\multirow{2}{*}{\textbf{Phase}}
 & \multicolumn{5}{c|}{\textbf{off-chip memory access}} & 
 
 \multirow{2}{*}{
 \begin{tabular}{c}
      \textbf{peak on-chip memory}  \\
      \textbf{requirement} 
 \end{tabular}
 } 
 \\ \cline{2-6}
 & \textbf{(I)NTT} & \textbf{LT matrix ($\underline{\hat{\mathbf{F}}}$)} & \textbf{switching key} & \textbf{polynomial read} & \textbf{polynomial write} & \\ \hline
 \textbf{1} & $3(L\!+\!1)+2\alpha$ & 0 & 
 \begin{tabular}{c}
      $2(n'_1\!-\!1)\beta$\\
      $(L \!+\! 1 \!+\! \alpha)$
 \end{tabular} 
 & $2(L \!+\! 1)$ & 
 \begin{tabular}{c}
      $3(n'_1-1)(L \!+\! 1 \!+\! \alpha)$ \\
      $+2(L + 1)$
 \end{tabular}
 &
 \begin{tabular}{c}
     $2(L+1)+(\beta+4)l_1$\\
     $+ (4\beta+6)m_1l_1$ 
 \end{tabular}
 \\ \hline
 \textbf{2} & $\frac{n'_1-1}{m_2}(L\!+\!1 \!+\! \alpha)$ & 0 & 0 & $(n'_1 \!-\! 1)(L \!+\! 1 \!+\! \alpha)$ & 
 \begin{tabular}{c}
       $(n'_1 \!\!-\!\! 1)\beta(L \!+\! 1 \!+\! \alpha)$ 
 \end{tabular}
 & 
 \begin{tabular}{c}
     $2(L + 1 + \alpha)+m_2(L+1$\\
     $+ 2\beta l_2) + 2l_2$ 
 \end{tabular}
 \\ \hline
 \textbf{3} & 0 & 0 & 
 \begin{tabular}{c}
       $2(n'_2 \!-\! 1)\beta$\\
       $(L \!+\! 1 \!+\! \alpha)$ 
 \end{tabular}
 & $\tfrac{n'_2-1}{m_3}n'_1(\beta \!+\! 1)(L \!+\! 1 \!+\! \alpha)$ & $2n'_1n'_2(L \!+\! 1 \!+\! \alpha)$ & 
 \begin{tabular}{c}
     $2l_3((\beta+1)m_4 + 2\beta m_3$\\
     $+2m_3m_4)$ 
 \end{tabular}
 \\ \hline
 \textbf{4} & $L\!+\!1\!+\!\alpha$ & $n(L\!+\!1\!+\!\alpha)$ & 0 &
 \begin{tabular}{c}
     $2(n'_1n'_2 \!\!+\!\! (\tfrac{n'_1n'_2}{m_5} \!\!-\!\! 1)n'_3)$ \\
     $(L\!+\!1\!+\!\alpha)$
 \end{tabular}  & 
 \begin{tabular}{c}
 $2\tfrac{n'_1n'_2}{m_5}n'_3$\\
 $(L\!+\!1\!+\!\alpha)$
 \end{tabular} 
 & 
 \begin{tabular}{c}
     $6m_5l_4 + 5l_4$
 \end{tabular}
 \\ \hline
 \textbf{5} 
 & 
 \begin{tabular}{c}
    $\tfrac{n'_3}{m_6}(L\!+\!1\!+\!\alpha)$ 
 \end{tabular}
 & 
 0 
 & 
 \begin{tabular}{c}
       $2(n'_3 \!-\! 1)\beta$\\
       $(L \!+\! 1 \!+\! \alpha)$ 
 \end{tabular}
 &
 \begin{tabular}{c}
     $2(n'_3 \!+\! \tfrac{n'_3}{m_6} \!-\! 1)$ \\
     $(L\!+\!1\!+\!\alpha)$
 \end{tabular}  
 & 
 \begin{tabular}{c}
 $2\tfrac{n'_3}{m_6}(L\!+\!1\!+\!\alpha)$
 \end{tabular} 
 & 
 \begin{tabular}{c}
     $m_6(L \!+\! 1 \!+\! \alpha) \!+\! $
     \\
     $(5\beta m_6 \!\!+\!\! 2m_6 \!\!+\!\! 6)l_5 \!+\! 2l_5$ 
 \end{tabular}
 \\ \hline
 \textbf{6} 
 & 
 \begin{tabular}{c}
    $L\!+\!1\!+\!\alpha$ 
 \end{tabular}
 & 
 0 
 & 
0
 &
 \begin{tabular}{c}
     $2(L\!+\!1\!+\!\alpha)$
 \end{tabular}  
 & 
 \begin{tabular}{c}
 $2(L\!+\!1)$
 \end{tabular} 
 & 
 \begin{tabular}{c}
     $2(L\!+\!1\!+\!\alpha)$ 
 \end{tabular}
 \\ \bottomrule
\end{tabular}
\end{table*}

\textbf{Phase 2}: This phase performs the computations in Lines 6--7 of Algorithm~\ref{alg: Proposed triple hoisted BSGS algorithm}, and its data flow is depicted in Fig.~\ref{fig: MO TH_BSGS BD}(b). A ping-pong buffer is utilized to load $\underline {\mathbf {b}}_i$ from off-chip memory. The result of the ModDown is stored in on-chip memory. Assume that each NTT in the Decompose is processing $l_2$ limbs simultaneously. If the Decompose is carried out on the ModDown result of $\underline {\mathbf {b}}_i$ one by one, then the twiddle factors of the NTTs need to be reloaded from off-chip memory for every $\underline {\mathbf {b}}_i$. To allow the tuning of on-chip memory requirement and off-chip memory access, assume that $m_2$ different $\underline {\mathbf {b}}_i$ are processed by ModDown before Decompose is carried out. In this case, the same $l_1$ limbs of each of the $m_2$ polynomials can share the same twiddle factors in the NTTs. The memory access summarized in Table \ref{tab: MO_TH_BSGS complexity} is based on these assumptions.

\textbf{Phase 3}: This phase executes Lines 8–11 of Algorithm \ref{alg: Proposed triple hoisted BSGS algorithm} and its data flow is shown in Fig. \ref{fig: MO TH_BSGS BD}(c). Each of $\underline{\mathbf{A}}_i$ $\underline{\mathbf{D}}_i$, and $\underline{\mathbf{SWK}}_{n'_1j}$ has $L+1+\alpha$, $\beta(L+1+\alpha)$, and $2\beta(L+1+\alpha)$ limbs. To minimize off-chip memory access, the maximum possible number of switching keys should be loaded into the available on-chip memory and reused as much as possible. Assume that $m_3$ ($1 \leq m_3 <n'_2$) switching keys can be loaded into on-chip memory, and the remaining on-chip memory can hold $m_4$ $\underline{\mathbf{A}}_i$ and $\underline{\mathbf{D}}_i$ with $l_3$ limbs each. Then the two loops in Lines 6-7 can be completed in $\lceil(L+1+\alpha)/l_3\rceil\lceil (n'_2-1)/m_3\rceil\lceil n'_1/m_4\rceil$ rounds.

\begin{table}[t]
\caption{Parallelisms for optimized datapath in the TH-BSGS algorithm.}
\label{tab: MO TH_BSGS parameters}
\centering

\renewcommand{\arraystretch}{1.1}
\setlength{\tabcolsep}{4pt}
\footnotesize

\begin{tabular}{c|p{0.72\columnwidth}}
\toprule
\textbf{Parameter} & \centering \textbf{Description} \arraybackslash \\
\midrule

$m_1$ & Number of switching keys processed simultaneously in Phase 1. \\

$m_2$ & Number of ModDown/Decompose operations performed simultaneously in Phase 2. \\

$m_3$ & Number of switching keys processed simultaneously in Phase 3. \\

$m_4$ & Number of $\underline{\mathbf{A}}_i$ and $\underline{\mathbf{D}}_i$ values processed simultaneously in Phase 3. \\

$m_5$ & Number of ciphertexts partially accumulated at a time in Phase 4. \\

$m_6$ & Number of ModDown/Decompose operations and switching keys processed simultaneously in Phase 5. \\

$l_p$ & Limb-level parallelism in Phase $1 \leq p \leq 5$. \\

\bottomrule
\end{tabular}
\end{table}

\textbf{Phase 4}: The computations in Lines 12-14 of Algorithm \ref{alg: Proposed triple hoisted BSGS algorithm} are realized in this phase, as shown in Fig. \ref{fig: MO TH_BSGS BD}(d). The polynomials $\underline{\mathbf{A}}_i$, $\underline{\mathbf{B}}_i$, $\underline{\mathbf{U}}_{0,k}$, and $\underline{\mathbf{U}}_{1,k}$ have the same number of limbs. However, the number of possible values of $k$ and $i$ are $n'_3$ and $n'_1n'_2$, respectively. Since $n'_3 \ll n'_1n'_2$, to minimize the off-chip memory access, $\underline{\mathbf{A}}_i$ and $\underline{\mathbf{B}}_i$ should be reused as much as possible once loaded into the on-chip memory. If the on-chip memory is not sufficient, intermediate polynomials $\underline{\mathbf{U}}_{0,k}$ and $\underline{\mathbf{U}}_{1,k}$ will be written/read to/from the off-chip memory. The number of polynomials and the number of limbs in each polynomial to process each time are set to $m_5$ and $l_4$, respectively, in our design. 

%\textcolor{red}{The limbs are iterated in the outermost loop, the $m_5$ polynomials processed at each iteration are traversed in the middle loop, and the index $k$ is iterated in the innermost loop. This loop ordering enables the reuse of the $l_4$ limbs of the $m_5$ polynomials $\underline{\mathbf{A}}_i$ and $\underline{\mathbf{B}}_i$, as well as the $l_4$ INTT twiddle factors, across the index $k$ under the limited on-chip memory. Since the polynomials $\underline{\mathbf{U}}_{0,k}$ and $\underline{\mathbf{U}}_{1,k}$ must be reloaded and updated for each group of $m_5$ polynomials $\underline{\mathbf{A}}_i$ and $\underline{\mathbf{B}}_i$ from the middle loop, $m_5$ is set to its maximum value to minimize off-chip memory accesses.}

\textbf{Phase 5}: Lines 15–18 of Algorithm \ref{alg: Proposed triple hoisted BSGS algorithm} are executed in this phase. The block diagram for this phase is illustrated in Fig. \ref{fig: MO TH_BSGS BD}(e). The number of polynomials and the number of limbs in each polynomial to process each time are set to $m_6$ and $l_5$, respectively. The design is similar to that for phases 1 and 2. Moreover, to minimize the reload of NTT twiddle factors and the polynomials $\underline{\mathbf{C}}_0$ and $\underline{\mathbf{C}}_1$, the parameter $m_6$ should be maximized.

\textbf{Phase 6}: This phase implements the combined ModDown and rescaling in Line 19 of Algorithm \ref{alg: Proposed triple hoisted BSGS algorithm} \cite{MAD}. All limbs in the polynomials are processed simultaneously. 
\begin{figure*}[t]
    \centering
    \includegraphics[width=1\linewidth]{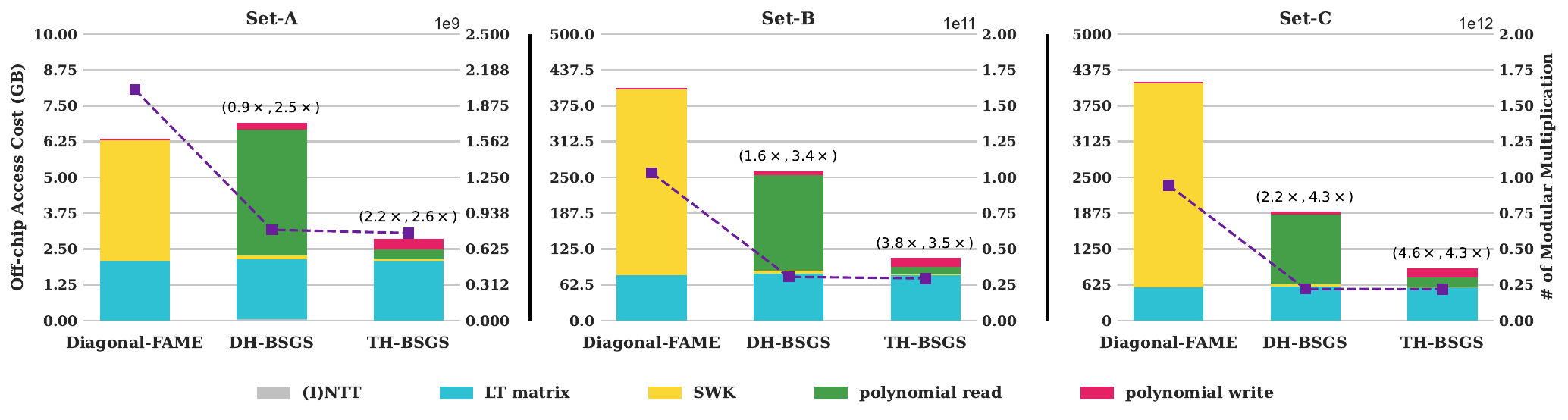}
    \vspace{-25pt}
    \caption{Comparisons of off-chip memory access (bar charts using the left axis) and number of modular multiplications (purple squares using the right axis) for the diagonal method \cite{LTHW-FAME}, DH-BSGS \cite{Bootstrapping4}, and proposed memory-optimized TH-BSGS method for HE parameters in Table \ref{tab: HE parameters}. The numbers inside parentheses represent the improvements in memory and computation complexity relative to the diagonal method \cite{LTHW-FAME}.}
    \label{fig: MO TH_BSGS Complexity Example}
    \vspace{-10pt}
\end{figure*}

The parallelisms utilized for the six phases are summarized in Table \ref{tab: MO TH_BSGS parameters}. According to the formulas provided in Table~\ref{tab: MO_TH_BSGS complexity}, the parallelism parameters listed in Table~\ref{tab: MO TH_BSGS parameters} can be tuned to minimize off-chip memory access and/or on-chip memory requirement.

\begin{table}[t]
\caption{Sets of HE parameters for 128-bit security used for evaluation.}
\label{tab: HE parameters}
\centering
\setlength{\tabcolsep}{7pt} % default is 6pt
\footnotesize
\begin{tabular}{c|ccccc}
\toprule
 & $N$ & $L+1$  & $\alpha$ & $\beta$ & $w$   \\ \hline
\textbf{Set-A} & $2^{13}$ & $5$ & $5$ & $1$ & $54$ 
\\ 
\textbf{Set-B} & $2^{15}$ & $16$ & $8$ & $2$ & $54$ 
\\ 
\textbf{Set-C} & $2^{16}$ & $32$ & $12$ & $3$ & $54$

% \\ \hline \hline
% & \textbf{Set-A0} & $2^{13}$ & $3$ & $1$ & $3$ & $54$ & $3.2$ & $128$ 
% \\ \hline
% & \textbf{Set-B0} & $2^{14}$ & $7$ & $1$ & $7$ & $54$ & $3.2$ & $128$ 
% \\ \hline
% & \textbf{Set-C0} & $2^{15}$ & $14$ & $1$ & $14$ & $54$ & $3.2$ & $128$ 
% \\ \hline \hline

% \textbf{Set-A1} & $2^{13}$ & $5$ & $5$ & $1$ & $54$ & $3.2$ & $128$ 
% \\ \hline
% \textbf{Set-B1} & $2^{15}$ & $16$ & $8$ & $2$ & $54$ & $3.2$ & $128$ 
% \\ \hline
% \textbf{Set-C1} & $2^{16}$ & $32$ & $12$ & $3$ & $54$ & $3.2$ & $128$ 
\\ \bottomrule
\end{tabular}
\end{table}

Consider the implementation on the Xilinx UltraScale+ U280 FPGA as an example. The device provides 43 MB of on-chip memory. Three sets of HE and LT parameters have been considered in \cite{LTHW-FAME} as listed in Table~\ref{tab: HE parameters}. For these sets of parameters, all feasible parallelism configurations that do not exceed the on-chip memory available are exploited, and the setting achieving the lowest off-chip memory access are plotted in Fig.~\ref{fig: MO TH_BSGS Complexity Example}. It can be observed that the proposed TH-BSGS design achieves more than 52\% reduction in off-chip memory access compared with the DH-BSGS scheme. The proposed scheme also significantly reduces the number of Decompose and ModDown operations required for ciphertext rotations, as shown in Table~\ref{tab: triple hoisted BSGS complexity analysis}. However, polynomial multiplications still account for the majority of the overall computational complexity. Consequently, the total number of modular multiplications required by the proposed design is only slightly lower than that of the DH-BSGS algorithm.

\section{Hardware Accelerator for HE-LT} \label{sec: Hardware Accelerator for HE-MVM}

\begin{figure}[t]
    \centering
    \includegraphics[width=0.95\linewidth]{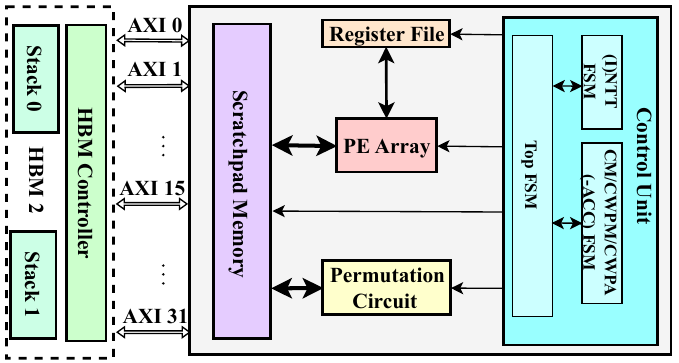}
    \vspace{-5pt}
    \caption{Top-level architecture of the proposed accelerator for HE-LT.}
    \label{fig: accelerator for HE-MVM}
\end{figure}

This section presents a hardware accelerator for the proposed memory-optimized TH-BSGS algorithm for HE-LT. In particular, a new permutation method is proposed to implement the automorphism in \eqref{eq: rotation automorphism function} without requiring extra memory buffers, while reducing the latency by a half. The top-level architecture of the proposed accelerator is illustrated in Fig.~\ref{fig: accelerator for HE-MVM}. It consists of five main components: an array of processing elements (PEs), a permutation circuit, a bank of scratchpad memories, a register file, and a control unit. The on-chip memory is the primary resource bottleneck of the HE-LT accelerator. Given the available on-chip memory blocks and the HE parameters, the number of coefficients in each polynomial that can be processed simultaneously and hence the number of PEs, denoted by $d_p$, is determined. Similar to the design in \cite{LTHW-FAME}, it is assumed that the off-chip memory is implemented using high-bandwidth memory (HBM) comprising two stacks, each with 16 AXI channels capable of transferring 256 bits of data in parallel. The details of each of the five components in the proposed accelerator are described below.

\begin{figure}
    \centering
    \includegraphics[width=0.95\linewidth]{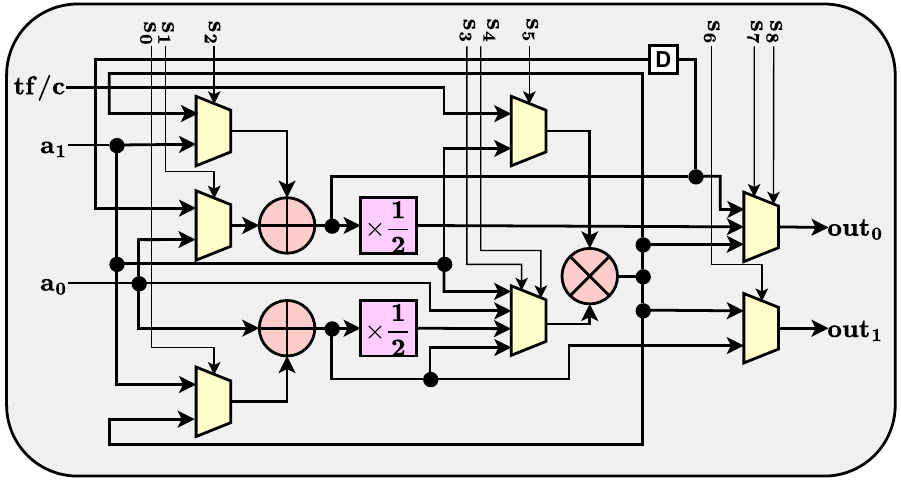}
    \vspace{-5pt}
    \caption{Architecture for the PE.}
    \label{fig: Design of the PE cores}
    \vspace{-10pt}
\end{figure}

\begin{table}[t]
\caption{PE control bits configuration for various operation modes.}
\label{tab: PE configuration}
\centering
\setlength{\tabcolsep}{8pt} % default is 6pt
\footnotesize
\begin{tabular}{c|ccccccc}
\toprule
 & $\mathbf{s_0}$ & $\mathbf{s_1}$ & $\mathbf{s_2}$ &  $\mathbf{s_3s_4}$ & $\mathbf{s_5}$ & $\mathbf{s_6}$ & $\mathbf{s_7s_8}$
\\ \hline
\textbf{NTT} & 1 & 1 & 0 & 00 & 0 & 1 & 00
\\
\textbf{INTT} & 0 & 1 & 1 & 01 & 0 & 0 & 10
\\
\textbf{CWPM} & x & x & x & 10 & 1 & x & 01
\\
\textbf{CWPA} & x & 1 & 1 & xx & x & x & 00
\\
\textbf{CM} & x & x & x & 10 & 0 & 0 & 01
\\
\textbf{CWPA-CM} & 0 & x & x & 11 & 0 & x & 01
\\
\textbf{CM-ACC} & x & 0 & 0 & 10 & 0 & x & 00
\\
\textbf{CWPM-ACC} & x & 0 & 0 & 10 & 1 & x & 00
\\ \bottomrule
\end{tabular}
\end{table}

\textbf{PE Array}: Each PE supports eight operating modes: butterfly operations for NTT and INTT, coefficient-wise polynomial multiplication (CWPM), coefficient-wise polynomial addition (CWPA), constant multiplication (CM), combined CWPA-CM operation, constant multiplication with accumulation (CM-ACC) for basis conversions in \eqref{eq: basis conversion}, and coefficient-wise polynomial multiplication with accumulation (CWPM-ACC) for inner product with switching keys and calculations in Lines 12 and 14 of Algorithm \ref{alg: Proposed triple hoisted BSGS algorithm}. In the NTT and INTT modes, each PE takes three inputs: $a_0$, $a_1$, and a twiddle factor $\mathrm{tf}$. In the NTT mode, the outputs are $a_0 + a_1 \cdot \mathrm{tf}$ and $a_0 - a_1 \cdot \mathrm{tf}$. In the INTT mode, the outputs are $(a_0 + a_1)/2$ and $((a_0 - a_1)/2)\cdot \mathrm{tf}$. Both of the CWPM and CWPA modes take $a_0$ and $a_1$ as the two inputs. Their outputs are $a_0 a_1$ and $a_0 + a_1$, respectively. In the CM mode, the inputs are a constant $c$ and a polynomial coefficient $a_0$, and the output is $c a_0$. In the CWPA-CM mode, the inputs are a constant $c$ and two polynomial coefficients, $a_0$ and $a_1$, and the output is $c(a_0 - a_1)$. The CM-ACC and CWPM-ACC accumulate the results from CM and CWPM computations, respectively. In every clock cycle, the output of the multiplier is accumulated to the intermediate value stored in a register. All the eight operating modes can be implemented using the PE architecture shown in Fig.~\ref{fig: Design of the PE cores} by configuring the multiplexer select signals as specified in Table~\ref{tab: PE configuration}. The adders, multiplier, and $\times \tfrac{1}{2}$ units perform modular arithmetic operations and can be implemented using the designs in \cite{ParhiNTT, MMIDSajjadZhangJSPS}. Wider multiplications can be broken down into the multiplications supported by the DSP blocks of FPGAs according to the design in \cite{FPGA-mod-mult}.

%and can be implemented by the DSP The modular multiplier is implemented using the FPGA’s DSP resources, following the design in \cite{FPGA-mod-mult}.

% \textbf{TA Array}: The number of inputs to a TA can be $\beta+1$, $\alpha$, or $m_5$, depending on the computation. To support the processing of $dp$ coefficients in each polynomial, $dp$ TAs are required for each polynomial. Accordingly, $n_{\text{TA}}$ can be determined based on the memory-optimized data path shown in Fig.~\ref{fig: MO TH_BSGS BD}.

%\begin{figure}[t]
 %   \centering
 %   \includegraphics[width=0.95\linewidth]{figures/multibanked_scratchpad_memory.drawio.pdf}
%    \vspace{-5pt}
%   \caption{Design of multi-banked scratchpad memory.}
%    \label{fig: Design of multi-banked scratchpad memory}
  %  \vspace{-10pt}
%\end{figure}

%\textbf{Multi-banked scratchpad memory}: The scratchpad memory in the proposed architecture is implemented using two types of on-chip memory resources: block RAM (BRAM) and UltraRAM (URAM) of the FPGA device. Each memory bank stores four types of data entities: intermediate ciphertext limbs, twiddle factors for (I)NTT operations, switching key limbs, and plaintext polynomials of the input matrix. The architecture of the scratchpad memory is illustrated in Fig.~\ref{fig: Design of multi-banked scratchpad memory}. During evaluation, each memory bank can transfer $dp$ coefficients or twiddle factors between the on-chip SRAM and the computation units. This enables the processing of $dp$ coefficients per limb in each cycle during the computations shown in Fig.~\ref{fig: MO TH_BSGS BD}.

\textbf{Permutation Circuit}: In prior designs \cite{LTHW-FAME, HE_MV_BWEfficient}, permutation circuits reorder polynomial coefficients to support two types of operations: (I)NTT and the $\phi_r(\cdot)$ automorphism in \eqref{eq: rotation automorphism function}. A circuit for supporting $d_p$-parallel permutation is proposed in \cite {SPN} and it has been utilized in \cite{LTHW-FAME, HE_MV_BWEfficient}. It consists of two $d_p$-input BENES networks, each of which consists of $d_p\log_2 (d_p)$ multiplexers, and an array of buffers along with its control units. It contributes to a large part of the overall HE-LT accelerator complexity and latency. The data in
each butterfly stage of the (I)NTT are accessed according to constant stride patterns \cite{ParhiNTT, Area-efficient-conflict-free}.
By exploiting these
switching patterns, the permutation circuit for (I)NTT can be eliminated by storing the polynomial coefficients in memory banks in row-major order and generating the memory addresses according to the access patterns. Although the output of the NTT in this case is in bit-reversed order, the following coefficient-wise computations and automorphism can be adjusted to continue in this order without any overhead. In the following, a simplified permutation network is developed for the automorphism by exploiting its mathematical formula. 

As shown in Fig.~\ref{fig: MO TH_BSGS BD}, the inputs to each $\phi_r(\cdot)$ function are in the NTT domain. In this case, the automorphism operation in \eqref{eq: rotation automorphism function} needs to be modified such that it maps the coefficient $A_{l}$ to $A_{l'}$, where
\begin{equation}\label{eq: auto}
l' = ((g_r(2l+1)\mod{2N})-1) /2,
\end{equation}
and $g_r = g^r \bmod 2N$ \cite{lattigoLibrary}. In our proposed design, the coefficients of an NTT-transformed polynomial are stored in $d_p$ memory blocks, where row $i$ of block $j$ stores the coefficient $A_{\bitrev(i\cdot d_p+j, \log_2 N)}$ ($0\leq i< N/d_p, 0\leq j<d_p $). The function $\bitrev(\cdot, n)$ reverses the bits of an $n$-bit integer. For example, $\bitrev(1,3)=\bitrev(\text{`001'},3)=\text{`100'}=4$. To carry out the automorphism, $d_p$ coefficients are read from row $i$ of the memory blocks. It will be shown next that the automorphism maps these $d_p$ coefficients to a set of $d_p$ coefficients whose indices are the same as those stored in row $i'$ of the $d_p$ memory blocks. In this case, the automorphism can be carried out next on row $i'$ of the memory blocks while the permuted row $i$ is written to memory address $i'$. Hence, no extra memory is needed to store the permuted result. Additionally, the permutation circuit is simplified utilizing the properties of the automorphism. 

The index of the coefficient stored in row $i$ and block $j$ of memory can be rewritten as 
\begin{align*}
   \bitrev(i\cdot d_p+j, \log_2 N) = b N/d_p + a,
\end{align*}
where $a = \bitrev(i, \log_2 (N/d_p))$ and $b= \bitrev(j, \log_2 d_p)$. It is assumed that $d_p$ is a power of two. Plugging in \eqref{eq: auto}, the automorphism maps this coefficient to the coefficient with index
 \begin{equation}\label{eq: perm 2}
    \begin{aligned}
       & \frac{\left[g_r\left(2bN/d_p + 2a + 1\right)\right]_{2N} - 1}{2} \\
        &= \left[\![g_rb]_{d_p}\tfrac{N}{d_p} + g_ra + \frac{g_r - 1}{2}\right]_N \\
        &= \left[g_rb + t_{i}\right]_{d_p}\frac{N}{d_p} + u_{i} = v_{j,i}\frac{N}{d_p} + u_i,
    \end{aligned}
 \end{equation}
 where $t_i$, and $u_i$ are the $\log_2 d_p$ most significant bits and $\log_2 (N/d_p)$ least significant bits in $[g_r a + (g_r - 1)/2]_N$, respectively. In another word, $[g_r a + (g_r - 1)/2]_N = t_i N/d_p + u_i$ for $0\leq t_i <d_p$ and $0\leq u_i < N/d_p$.

Consider two coefficients from the same row $i$ but two distinct memory blocks $j_1$ and $j_2 \neq j_1$. Their $u_i$ and $t_i$ are the same. However, $v_{j_1,i} \neq v_{j_2,i}$. This can be proved by contradiction. If $v_{j_1,i} = v_{j_2,i}$, then $[g_r b_1]_{d_p} = [g_r b_2]_{d_p}$. Since $g_r$ is odd and $d_p$ is a power of two, $g_r$ is invertible modulo $d_p$. Therefore, 
$b_1 = b_2$.
Because $b_1=\bitrev(j_1,\log d_p)$ and $b_2=\bitrev(j_2,\log d_p)$, this gives $\bitrev(j_1,\log d_p)=\bitrev(j_2,\log d_p)$, which contradicts $j_1 \neq j_2$. Hence, $v_{j_1,i} \neq v_{j_2,i}$. Considering that all the coefficients are stored in bit-reverse order in the memory, the polynomial coefficients from row $i$ of the memory blocks become the coefficients with indices   
\begin{align*}
    \bitrev(v_{j,i}N/d_p + u_i, \log N) = i^\prime d_p + j'
\end{align*}
after the automorphism. Here  $i^\prime = \bitrev(u_i, \log N/d_p)$ consists of the higher MSBs and $j' = \bitrev(v_{j,i}, \log d_p)$ has the lower LSBs. These coefficients
are written to row $i'$ of the memory blocks. 

In each clock cycle, $d_p$ coefficients from row $i'$ of the memory blocks are fetched, and the automorphism results of the coefficients from row $i$ are written to row $i'$. However, starting from a certain row $i$, such a process may go back to row $i$ before every of the $N/d_p$ rows of the memory blocks are covered. To solve this problem, a flag is introduced for each memory row. Once a row has been processed, the corresponding flag is set. If row $i'$ has a set flag, then the next row without a set flag is selected and processed. Such a selection can be implemented by a 2-layer priority encoder with low complexity. 

From the above analysis, our design only needs a permutation circuit that switches the $d_p$ inputs to the $d_p$ outputs, which can be implemented by a $d_p$-input BENES network that has $d_p\log{d_p}$ multiplexers. These multiplexers are controlled by $j^\prime$, which can either be computed on-the-fly or precomputed and stored in a look-up table (LUT). Compared with the permutation circuit in \cite{SPN}, our proposed design requires only one instead of two BENES networks, thereby reducing the latency by half. Our design also eliminates the additional buffers. 

\begin{figure}[t]
    \centering
    \includegraphics[width=0.9\linewidth]{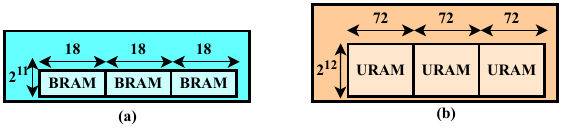}
    \caption {Design of the memory block: (a) memory block $2K \times 54$ using BRAMs; (b) memory block $4K \times 4 \times 54$ using URAMs.
    }
    \label{fig: memory banks}
\end{figure}

\textbf{Register File}: The register file in the proposed design stores the moduli $q_j$, $p_i$ and their related values. Moreover, during the (I)NTT operation, the $d_p$ PEs need the polynomial coefficients stored in two different rows of the memory blocks and generate two sets of output coefficients. The inputs and outputs of the PEs need to be held temporarily in the register file after (before) they are read from (written back) to the memory blocks. 

\textbf{Scratchpad Memory}: Similar to prior designs \cite{LTHW-FAME}, our accelerator utilizes $w$=54 bits to represent each polynomial coefficients. The width of the BRAM and URAM on FPGA devices may not match this bit width exactly. Multiple BRAMs and URAMs are put together to implement memory blocks needed in our design. For example, the width of the BRAMs on the Xilinx U280 device is 18 bits. Three of them are put together in our design to implement a memory block holding polynomial coefficients as shown in Fig. \ref{fig: memory banks}(a). URAMs have a width of 72 bits. Since the polynomial coefficients for different limbs are accessed in the same order, three copies of URAMs are put together to implement $3\times 72/54=4$ memory blocks as illustrated in Fig. \ref{fig: memory banks}(b). Memory blocks with the same data access pattern are grouped to share the same read/write addresses, allowing for reading $d_p$ coefficients simultaneously. In the case that the depth of the BRAM/URAM is sufficient to hold multiple polynomials, the polynomials that are not accessed simultaneously are stored in the same BRAM/URAM to more efficiently utilize the on-chip memory resources.

\textbf{Control Unit}: The control unit generates the read/write addresses for the memory blocks and all necessary control signals. It consists of an (I)NTT finite state machine (FSM), a CWPM/PA/PA-PM/CM(-ACC) FSM, and a top FSM. The (I)NTT FSM generates the read/write addresses of the coefficients and twiddle factors needed for each clock cycle of the (I)NTT operations. The CWPM/PA/PA-PM/CM(-ACC) FSM generates the read/write addresses for the CWPM, CWPA, CWPA-PM, CM, CWPM-ACC, and CM-ACC operation modes used in Decompose, ModUp, ModDown, basis conversion, inner product with switching keys, and plaintext-ciphertext multiplication in the LT. The top FSM translates these virtual addresses into their corresponding physical addresses for the on-chip memories and generates all necessary control signals to implement the memory-optimized data path shown in Fig. \ref{fig: MO TH_BSGS BD}.

\begin{table*}[t]
\caption{The algorithmic and hardware parameters for different HE parameter sets.}
\label{tab: MO_TH_BSGS parameters}
\centering
\setlength{\tabcolsep}{6pt} % default is 6pt
\footnotesize
\begin{tabular}{c|ccccccc|cccc}
\toprule
& $\mathbf{n}$ & $\mathbf{(n'_1,n'_2,n'_3)}$ & $\mathbf{(m_1,l_1)}$ & $\mathbf{(m_2,l_2)}$ & $\mathbf{(m_4, m_3, l_3)}$ & $\mathbf{(m_5,l_4)}$ & $\mathbf{(m_6,l_5)}$ & $\mathbf{d_p}$ 
\\ \hline
\textbf{Set-A} & $2^{12}$ & $(2^4,2^4,2^4)$ & $(5,5)$ & $(15,10)$ & $(8, 15, 1)$ & $(86,1)$ & $(8,4)$ & $128$ 
\\ 
\textbf{Set-B} & $2^{14}$ & $(2^5,2^4,2^5)$ & $(8,1)$ & $(1,12)$ & $(1,8,1)$ & $(25,1)$ & $(4,1)$ & $128$ 
\\ 
\textbf{Set-C} & $2^{15}$ & $(2^5,2^5,2^5)$ & $(1,1)$ & $(1,1)$ & $(1,4,1)$ & $(12, 1)$ & $(1,1)$ & $256$ 
\\ \bottomrule
\end{tabular}
\end{table*}

\begin{figure*}[t]
    \centering
    \includegraphics[width=1\linewidth]{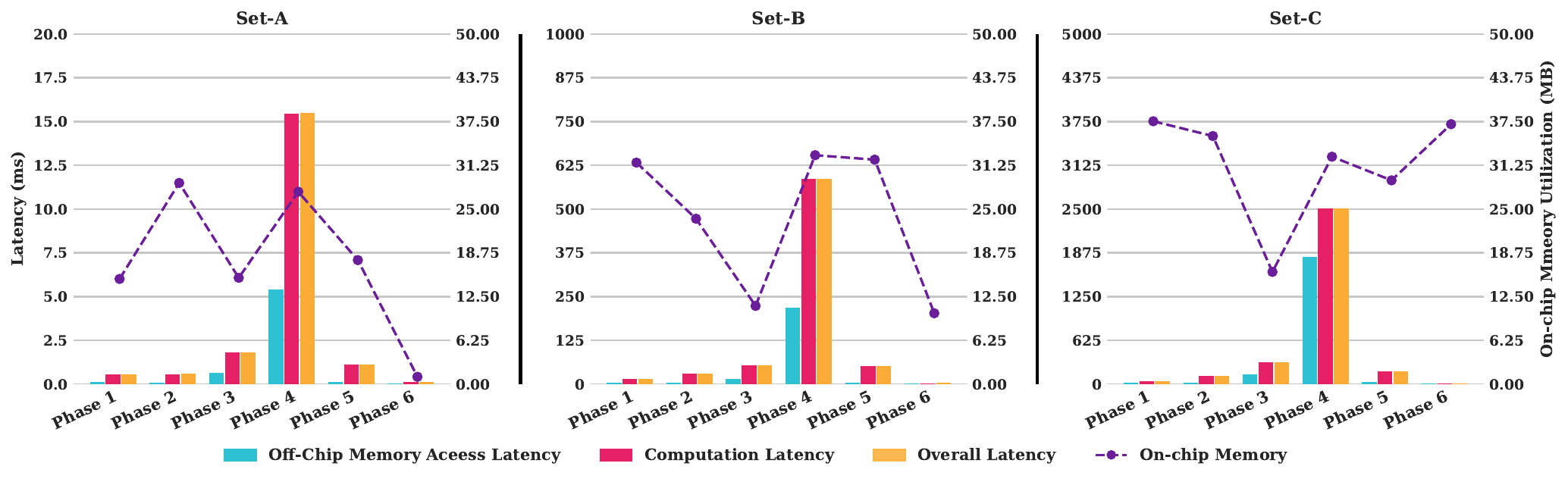}
    \vspace{-25pt}
    \caption{Off-chip memory access latency, computation latency, end-to-end latency, and on-chip memory requirement across the six phases of the proposed memory-optimized TH-BSGS method for the three HE parameter sets listed in Table~\ref{tab: HE parameters}.}
    \label{fig: latency analysis}
    \vspace{-10pt}
\end{figure*}

\section{Experimental Results and Comparisons}
This section first presents the experimental evaluation setup on FPGAs. Then the performance of our design is analyzed and compared with those of prior efforts.  

\begin{table*}[t]
\caption{The hardware resource utilization of the proposed work compared with those of prior HE-LT accelerators.}
\label{tab: HW resource utilization}
\centering
\setlength{\tabcolsep}{4.5pt} % default is 6pt
\footnotesize
\begin{threeparttable}
\begin{tabular}{c|c|cccccc|cc | c} 
\toprule
\textbf{Design} & \textbf{Parameters} & \textbf{Board} & \textbf{kLUT} & \textbf{kFF} & \textbf{BRAM} & \textbf{URAM} & \textbf{DSP} & \textbf{Freq. (MHz)} & \begin{tabular}{c}
     \textbf{Overall} \\
     \textbf{Latency (ms)}
\end{tabular}
&
\begin{tabular}{c}
     \textbf{Off-Chip Mem.} \\
     \textbf{Access (GB)}
\end{tabular}
\\ \hline
\multirow{3}{*}{Design in \cite{HE_MV_BWEfficient}}& Set-A$'$ & U200 & 756.3 & 610.4 & 224 & 656 & 4896 & 200 & 773.96 & 5.70
\\ 
& Set-B$'$ & U200 & 835.6 & 1093.2 & 896 & 742 & 5712 & 180 & 11450.86 & 100.40
\\ 
& Set-C$'$ & U200 & 835.6 & 1093.2 & 896 & 742 & 5712 & 180 &  154749.49 & 1464.68
\\ \hline
\multirow{3}{*}{FAME \cite{LTHW-FAME}}& Set-A
% \tnote{$\dagger$} 
& U280 & 636.0 & 998.0 & 1024 & 0 & 5376 & 350 & 54.17 & 6.33
\\ 
& Set-B & U280 & 701.0 & 1147.0 & 640 & 192 & 5376 & 350 & 2968.53 & 405.02
\\ 
& Set-C & U280 & 803.0 & 1660.0 & 3328 & 672 & 5376 & 300 & 17616.38 & 4158.02
\\ \hline
\multirow{3}{*}{\begin{tabular}{c}
     DH-BSGS  \cite{Bootstrapping4}
\end{tabular}}& Set-A & U280 & 303.1  & 216.5 & 1536 & 672 & 2176 & 350 & 36.58 & 6.87
\\ 
& Set-B & U280 & 303.9 & 216.7 & 1536 & 768 & 2176 & 350 & 1297.14 & 258.81
\\ 
& Set-C & U280 & 617.8 & 432.8 & 1920 & 768 & 4352 &  300 & 7494.10 & 1885.85
\\ \hline
\multirow{3}{*}{Proposed}& Set-A & U280 & 303.1  & 216.5 & 1536 & 672 & 2176 & 350 & 19.44 & 2.84
\\ 
& Set-B & U280 & 303.9 & 216.7 & 1536 & 768 & 2176 & 350 & 726.59 & 107.91
\\ 
& Set-C & U280 & 617.8 & 432.8 & 1920 & 768 & 4352 &  300 & 3113.70 & 903.47
\\ \bottomrule
\end{tabular}
\begin{tablenotes}
\footnotesize
\item[$^*$] The Set-A$'$, B$'$, and C$'$ parameters used in \cite{HE_MV_BWEfficient} are smaller than those in Set-A, B, and C listed in Table \ref{tab: HE parameters}. 
\end{tablenotes}
\end{threeparttable}
\end{table*}

% \subsection{Experimental Setup}
The proposed HE-LT accelerator is described in System Verilog HDL and synthesized over the Alveo U280 board with Virtex UltraScale+ XCU280 FPGA using Vivado 2023.2. This FPGA contains 1304K LUTs, 2607K FFs, 2016 BRAMs, 960 URAMs, and 9024 DSPs. The proposed design is synthesized for the three sets of HE parameters listed in Table~\ref{tab: HE parameters}, each supporting a different multiplication depth, $L$, for various applications. The same parameters were used in prior design \cite{LTHW-FAME}. For HE defined over the ring $\mathcal{R}_Q = \mathbb{Z}_Q[x]/(x^N+1)$, at most $N/2$ slots can be packed into a ciphertext. Therefore, an LT dimension of $n = N/2$ is considered in the evaluation, as listed in Table~\ref{tab: MO_TH_BSGS parameters}. The values of $n'_1$, $n'_2$, and $n'_3$ listed in this table minimize the computation complexity. Given the amount of on-chip memory available on the XCU280 FPGA, a compiler is developed to determine the maximum parallelism that can be adopted in each phase. For the parameters listed in Table~\ref{tab: MO_TH_BSGS complexity}, the maximum allowed parallelism are listed in Table \ref{tab: MO_TH_BSGS parameters}. The number of coefficients from each polynomial that can be processed simultaneously, $d_p$, is determined by the B/URAM size and the value of $N$. 

The FPGA resource utilization obtained from synthesizing under the same clock frequency used in \cite{LTHW-FAME} are reported in Table~\ref{tab: HW resource utilization}. Since a cycle-accurate simulation model for off-chip memory access is not available, the latency of off-chip memory access is estimated by dividing the amount of data to access by the bandwidth, which is 480GB/s for the U280 board. The computation latency can be calculated by multiplying the number of clock cycles with the clock period. These two latencies for each phase are shown in Fig. \ref{fig: latency analysis}. The off-chip memory access and computation can be overlapped except for the loading of the first set of data and writing back of the last set of data. Hence, the end-to-end latency of each phase is almost equal to the maximum of the off-chip memory access latency and the computation latency. As it was shown in Fig. \ref{fig: MO TH_BSGS Complexity Example}, our optimized data path reduces the off-chip memory access by a significant portion. Hence, the off-chip memory access no longer dominates the latency of each phase when the bandwidth is high. Our proposed design also substantially reduces the number of Decompose, ModDown, and switching key multiplication operations. As a result, Phase 4 for polynomial multiplication is now dominating the latency. Adding up the end-to-end latency of each phase, the overall latency of the HE-LT can be derived as listed in Table \ref{tab: HW resource utilization}. 

The design in \cite{HE_MV_BWEfficient} and FAME \cite{LTHW-FAME} are among the most efficient hardware accelerators for HE-LT. Both of them are based on  the diagonal method and their synthesis results are listed in Table \ref{tab: HW resource utilization} for comparison. Despite the smaller parameters used, the design in \cite{HE_MV_BWEfficient} has very long latency. 
FAME uses the same set of HE parameters as our proposed design. Compared to FAME with the Set-C parameters, our design reduces the overall HE-LT latency and off-chip memory access by 5.6$\times$ and 4.6$\times$, respectively. Besides, our design reduces the numbers of LUTs, flip-flops (FFs), and DSP blocks by 23\%,  74\%, and 19\%, respectively, and utilizes less BRAM and URAM.   

The DH-BSGS algorithm from \cite{Bootstrapping4} has not been implemented by hardware in any existing literature. The same hardware for our proposed TH-BSGS algorithm can be utilized to implement the DH-BSGS algorithm according to the data flow in Algorithm \ref{alg: double hoisting bsgs}, and the results are listed in Table \ref{tab: HW resource utilization}. Without our data path optimization proposed in Section \ref{sec: memory-optimized MVM datapath}, the large number of intermediate results in the DH-BSGS algorithm do not fit into on-chip memory and need to be loaded into and read from off-chip memory. First of all, this leads to more than twice off-chip memory access compared to our proposed TH-BSGS implementation. Secondly, the computations have to wait until the intermediate results are read from off-chip memory, and hence the computation latencies are not overlapped with the off-chip memory access latency. As a result, the overall latency of the DH-BSGS implementation is much longer than that of our proposed TH-BSGS implementation. For the Set-C parameters, its latency is 2.4$\times$ longer. 

\begin{table}[t]
\caption{Comparison of TH-BSGS HE-LT hardware accelerators with different permutation circuits}
\label{tab: PC performance}
\centering
\setlength{\tabcolsep}{5pt}
\footnotesize
\begin{tabular}{c|cc|cc}
\toprule
\multirow{3}{*}{\textbf{HE Parameter}}
& \multicolumn{2}{c|}{\textbf{Perm. Circ. in \cite{SPN}}}
& \multicolumn{2}{c}{\textbf{Prop. Perm. Circ.}} \\
% \cline{2-5}
\cmidrule(lr){2-3}
\cmidrule(lr){4-5}
& \textbf{Overall} & \multirow{2}{*}{\textbf{kLUT}}
& \textbf{Overall} & \multirow{2}{*}{\textbf{kLUT}} \\
& \textbf{Latency (ms)} & 
& \textbf{Latency (ms)} &  \\
\midrule
Set-A & 25.60   & 411.4  & 19.44 & 303.1 \\
Set-B & 891.11  & 411.4 & 726.59 & 303.9 \\
Set-C & 3650.84 & 845.0 & 3113.70 & 617.8 \\
\bottomrule
\end{tabular}
\end{table}

To show the improvements brought by our proposed permutation circuit, the TH-BSGS HE-LT accelerator using the permutation circuit from \cite{SPN} is synthesized, and the comparisons are listed in 
Table~\ref{tab: PC performance}. Since our design eliminates the permutations within (I)NTT operations and uses one instead of two BENES networks for the permutation needed in the automorphism, the overall latency is reduced by at least 15\% over the three sets of parameters. The reduced number of BENES networks also helps to reduce the number of LUTs needed on the FPGA devices by 27\%. The buffers in the permutation circuit of \cite{SPN} are implemented by sharing BRAMs. Hence, the resource requirements for BRAMs, FFs, and DSPs are the same and are not listed in Table \ref{tab: PC performance}. 

The improvements achieved by the proposed TH-BSGS accelerator depend on the off-chip memory bandwidth, the LT dimension $n$, and the decomposition factor $\beta$. The overall latency of the DH-BSGS algorithm is the computation latency plus the off-chip memory access latency. On the other hand, the overall latency of the proposed TH-BSGS design is the maximum of the memory access latency and computation latency. Hence, regardless of the memory bandwidth, which changes the memory access latency, our TH-BSGS design can achieve significant latency reduction. The FAME design also allows the computation and memory access latencies to be overlapped. However, since our design has much simpler computation and a smaller amount of off-chip data access, it can always achieve significant latency saving compared to FAME for a different bandwidth. The major contributors of computation complexity and off-chip memory access are plaintext-ciphertext multiplications and rotations. The rotation complexity scales as $\beta n$, $\beta\sqrt{n}$, and $\beta\sqrt[3]{n}$ for FAME, DH-BSGS, and the proposed TH-BSGS, respectively, while the plaintext–ciphertext multiplication complexity scales linearly with $n$ for all three designs. Hence, as $\beta$ increases, the achievable savings of our proposed design over prior efforts become more significant. On the other hand, as $n$ reduces, the proposed design has less significant improvement. 

\section{Conclusions}
This paper proposes a TH-BSGS algorithm for HE-LT. The decomposition of the baby step enables reformulations that eliminate redundant computations and combine complex operations efficiently. Furthermore, an optimized data path is proposed to maximize on-chip data reuse and reduce off-chip memory accesses. An efficient hardware accelerator is also developed for the proposed algorithm. By exploiting the automorphism pattern, the permutation circuit is significantly simplified. Compared with prior hardware accelerators, the proposed design substantially reduces both latency and off-chip memory access while requiring fewer hardware resources. Future work will focus on further optimizing linear transformations for specific applications.

% \begin{figure}
%     \centering
%     \includegraphics[width=0.95\linewidth]{figures/latency_breakdown_HE_MVM.pdf}
%     \caption{The latency breakdown of the proposed HE-MVM accelerator for the three set of parameters in Table \ref{tab: HE parameters}.}
%     \label{fig: latency breakdown}
% \end{figure}

% Fig. \ref{fig: latency breakdown} illustrates the latency breakdown of the various computation components in the proposed HE-MVM accelerator for the three parameter sets listed in Table \ref{tab: HE parameters}. As can be seen, the (I)NTT operations and the pt-ct multiplications with the diagonal plaintext polynomials of the input ciphertext account for the majority of the overall latency. The (I)NTT operations have significantly higher complexity than the other computation units, and the pt-ct multiplications are repeated $n$ times according to Algorithm \ref{alg: Proposed triple hoisted BSGS algorithm}, whereas the other computations are repeated only $n'_1$, $n'_2$, and $n'_3$ times. In addition, the (I)NTT operations are also involved in the Decompose and ModDown operations, further increasing their contribution and making them the main computational bottleneck during evaluation.

\printbibliography

\begin{IEEEbiography}[{\includegraphics[width=1in,height=1.25in, clip,keepaspectratio]{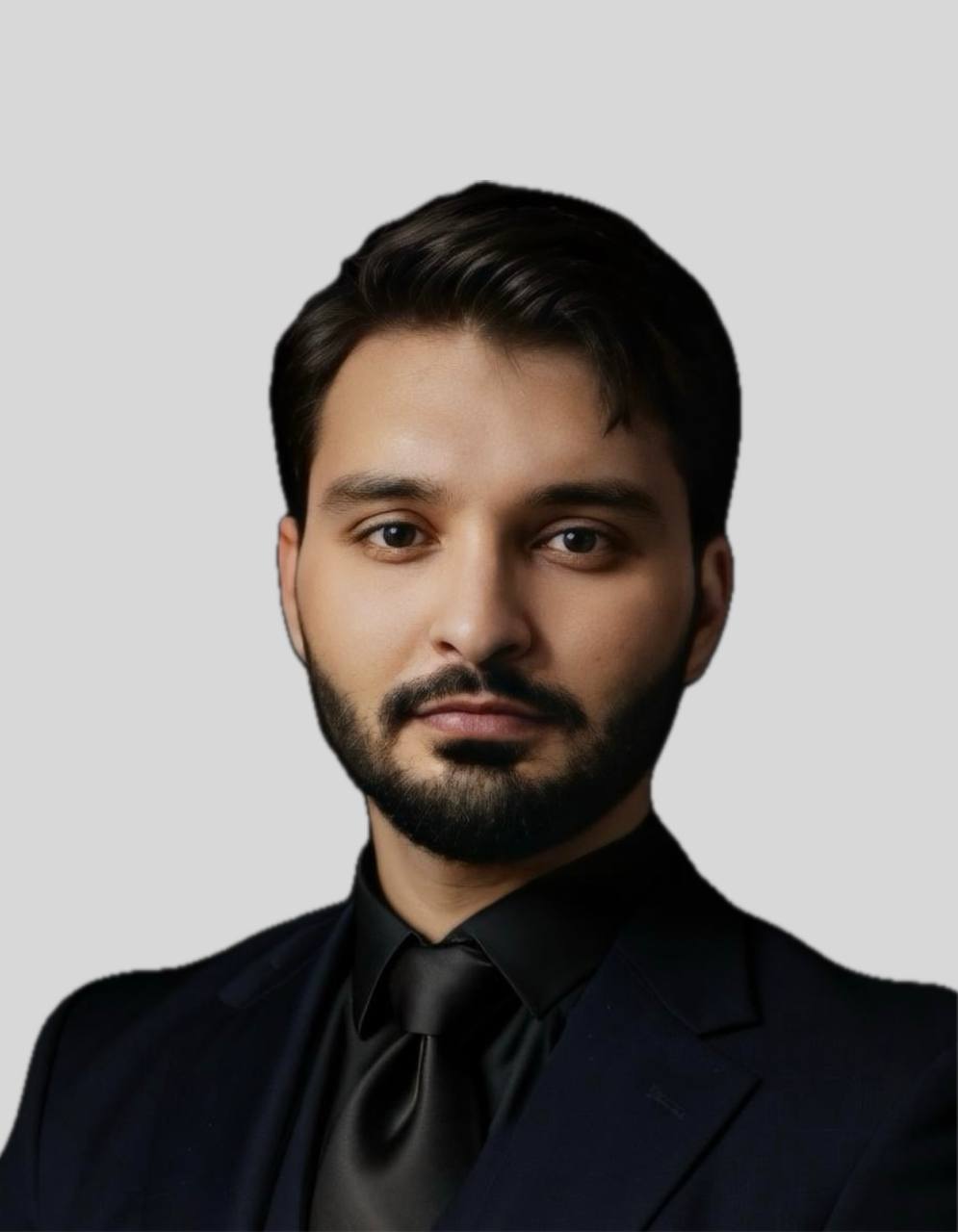}}]
{Sajjad Akherati} (Graduate Student Member, IEEE) received the B.Sc. degree in electrical engineering from Sharif University of Technology, Iran, in 2022. He is currently pursuing his Ph.D. degree with the Electrical and Computer Engineering Department, The Ohio State University, Columbus, OH, USA. His current research interests include VLSI architectures design for Homomorphic Encryption.
\end{IEEEbiography}

\begin{IEEEbiography}[{\includegraphics[width=1in,height=1.25in, clip,keepaspectratio]{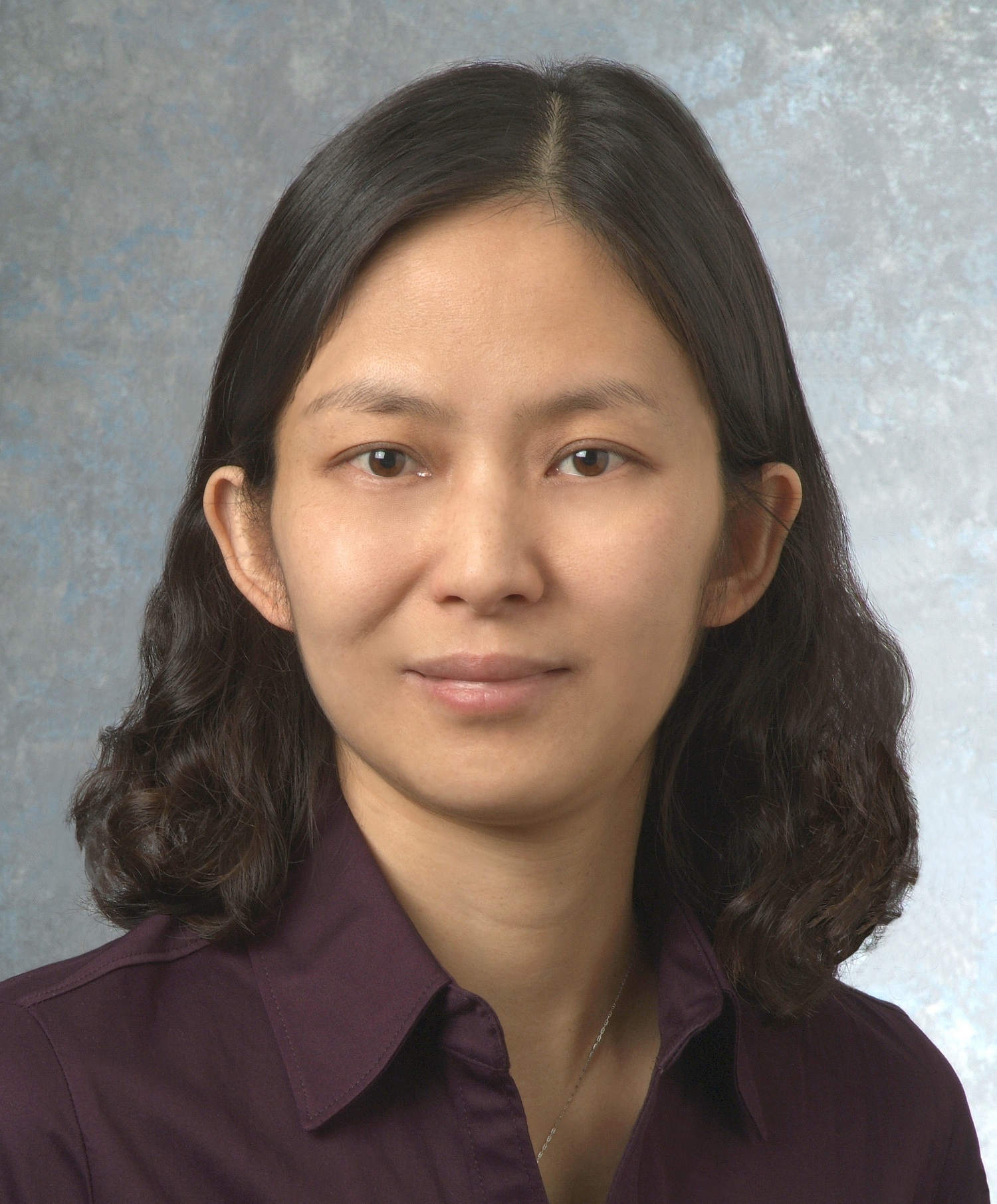}}]
{Xinmiao Zhang} (Fellow, IEEE) received her Ph.D. degree in Electrical Engineering from the University of Minnesota. She is currently a Professor at the Ohio State University. Prof. Zhang’s research spans the areas of VLSI architecture design, digital storage and communications, cryptography, security, and signal processing. Prof. Zhang is a recipient of the NSF CAREER Award 2009, and the College of Engineering Lumley Research Award at The Ohio State University 2022. Prof. Zhang was elected the Vice President-Technical Activities of the IEEE Circuits and Systems Society (CASS) 2022-2023 and served on the Board of Governors of CASS 2019-2021. She also served on the technical program and organization committees of many conferences, including ISCAS, ICC, GLOBECOM, SiPS, GlobalSIP, MWSCAS, and GLSVLSI. She is the Associate Editor-in-Chief for the IEEE Transactions on Circuits and Systems-I 2024-2027.
\end{IEEEbiography}

\end{document}